\documentclass[11pt]{article}
\usepackage{epsfig}

\newcommand{\be}{\begin{eqnarray}}
\newcommand{\ee}{\end{eqnarray}}

\def\bd{\begin{displaymath}}
\def\ed{\end{displaymath}}

\def\ba#1{\begin{array}{#1}}
\def\ea{\end{array}}
\def\nn{\nonumber}
\newfont{\Bbb}{msbm10 scaled 1200}

\begin{document}

\pagestyle{empty}

\begin{center}

{\LARGE\bf  Fundamental solution method 
   applied to time evolution of two energy level systems:\\[0.5cm]
exact and adiabatic limit results}
\vskip 18pt

{\large {\bf Stefan Giller{$\dag$} and Cezary Gonera{$\ddag$}}}

\vskip 3pt

Theoretical Physics Department II, University of {\L}\'od\'z,\\
Pomorska 149/153, 90-236 {\L}\'od\'z, Poland\\ 
e-mail: $\dag$ sgiller@krysia.uni.lodz.pl \\ 
$\ddag$ cgonera@krysia.uni.lodz.pl
\end{center}
\vspace{6pt} 
\begin{abstract}A method of fundamental solutions has been used to investigate 
     transitions in two energy level systems with no level crossing in a real time. Compact formulas for transition probabilities have been found in their exact form as
     well as in their adiabatic limit. No interference effects resulting from many level complex
     crossings as announced by Joye, Mileti and Pfister
     (Phys. Rev. {\bf A44} 4280 (1991)) have
     been detected in either case. It is argued that these results of this
     work are incorrect. However, some effects of Berry's phases are confirmed.

\end{abstract}
\vskip 9pt

{\small PACS number(s): 03.65.-W , 03.65.Sq , 02.30.Lt , 02.30.Mv}

{\small Key Words: two energy level systems, fundamental solutions, semiclassical expansion, adiabatic approximation}

\newpage

\pagestyle{plain}

\setcounter{page}{1}

\section*{1.  Introduction}

\hskip+2em Transitions between energy levels in a two energy level
 system evolving in time
 are of great importance
from many points of view. On one side such systems provide us with the simplest models
to investigate transition amplitudes between
different energy levels by different approaches \cite{1}. On the other side these systems play an important role in
experimental investigations of basic principles of quantum
 mechanics \cite{2}. Recently a lot of effort has been devoted to obtain more rigorous results on
the adiabatic limit of transition amplitudes for these systems
 \cite{3,4,5,6,7}. In particular in a series
of recent papers Joye \underline{et al} have studied this problem by the Hilbert space methods. Such
two energy level systems are formally equivalent to a one-half spin system put into time
dependent magnetic field. However  good approximate results and the more so the exact ones are difficult to
obtain for such systems even for simple time evolutions of the effective 'magnetic'
field. Therefore each opportunity of improving this situation is worth trying. 
A treatment of the problem by a method of fundamental solutions (so fruitful in its application
to stationary problems of 1-dim Schr\"odinger equation \cite{8,9,10}) is of first importance, the more
so that to our knowledge, the method was not used so far to
 this goal. A possibility
of application of the method is related to the fact that a linear system of first order
differential equations describing time evolution of transition
 amplitudes can always be transformed into a system of decoupled second order
equations having a form of the stationary Schr\"odinger equation, one for each amplitude. This allows us to apply all advantages of the fundamental solution method \cite{10}. The
only obstacle related with this approach is a complexity of effective 'potentials' which appear
in the final system of the Schr\"odinger-type equations. 

     The paper is organized as follows.

     In the next section the problem of transitions in two energy level systems is stated
and corresponding assumptions about the effective 'magnetic field' are formulated. A linear
system of two differential equations for the transition amplitudes is rewritten
in a form of two decoupled equations of the Schr\"odinger type. 

     In Sec. 3 properties of the fundamental solution method are recalled.

     In Sec.4 some subtleties of the application of the fundamental solution method to the
problems considered in the paper are discussed.

     The method is first applied to a particular system of the atom - atom scattering
within a frame of the Nikitin model \cite{11,12} in Sec.5 .

     In Sec.6 results of Sec.5 are next generalized to systems with an algebraic time
dependence of the effective magnetic field.

     In Sec.7 another two examples of two energy level systems are considered
with corresponding magnetic fields depending exponentially on time. These examples show
that a way the magnetic fields depend on time does not affect a form of
 the transition amplitudes. This form is not affected either by the number of (complex) energy level 
crossings on the Stokes lines
closest to the real axis of the complex time plane. The latter result confirms the one of the
previous section. Such a dependence resulting with some interference effects has been announced
by Joye \underline{et al} \cite{4}. 

In Sec.8 we consider an example of the magnetic field with an
     explicit contribution of the geometrical (Berry) phase to the
     transition probability.

     We summarize and discuss our results in the last section . In particular we show there 
     that the results of Joye, Mileti and Pfister [4] on the effects of
     interference from many level crossings are incorrect.

\section*{2.  Adiabatic transitions in two energy level systems}

\hskip+2em In general, any two energy level system is formally equivalent to a
one-half spin system put into an external magnetic field ${\bf
  B}(t)$. Therefore, we shall consider just such a system. Its
Hamiltonian $H(t)$ is given then by $H(t)=\frac{1}{2}\mu {\bf
  B}(t)\cdot{\bf \sigma}$ , where ${\bf \sigma}=(\sigma_x,\sigma_y,\sigma_z)$ are Pauli's matrices so
that two energy levels $E_{\pm}(t)$ of $H(t)$ are given by 
$E_{\pm}(t)=\pm \frac{\mu}{2}B(t)$ where $B(t)=\sqrt{{\bf B}^2(t)}$. 

When the adiabatic transitions between the two energy levels $E_{\pm}(t)$
 are considered then
the following properties of the field ${\bf B}(t)$ are typically
 assumed to be:

${\bf 1}^0$ ${\bf B}(t)$ is real being defined for the real $t$, $-\infty  <t<+\infty$;

${\bf 2}^0$ ${\bf B}(t)$ can be continued analytically off the real values of $t$ 
as a meromorphic function
defined on some $t$-Riemann surface ${\bf R}_B$. A sheet of ${\bf R}_B$
 from which ${\bf B}(t)$ is originally     continued is called physical; 

${\bf 3}^0$ On the physical sheet ${\bf B}(t)$ is analytic in an infinite strip
 $\Sigma=\{t:\vert \Im t \vert<\delta, \delta>0\}$,
 without roots in the strip and achieves there finite limits for 
$\Re t=\pm \infty$ , i.e.  ${\bf  B}(\Re t=\pm \infty)=
{\bf B}^\pm \ne{\bf 0}$ in the strip; 

    The field ${\bf B}(t)$ depends additionally on a parameter $T (>0)$
 i.e. ${\bf B}(t)\equiv{\bf B}(t,T)$ which
introduces a "natural" scale of time to the system, so that its time
 evolution is
expressed most naturally in units of $T$. If $T$ is small in comparison with 
the actual period of the
process considered then the latter is "fast" or "sudden". If, 
however, $T$ is large in
this comparison then the process is "slow" or "adiabatic". 

In the adiabatic process of the system the following is
  assumed about ${\bf B}(t,T)$:
     
${\bf 4}^0$ A dependence of ${\bf B}(t,T)$ on $T$ is such that a rescaled 
field ${\bf B}(sT,T)$ has the
     following asymptotic behavior for $T\to +\infty$ 
\begin{eqnarray}
{\bf B}(sT,T)\sim{\bf B}_0(s)+\frac{1}{T}{\bf B}_1(s)+
\frac{1}{T^2}{\bf B}_2(s)+\dots
\label{2.1}
\end{eqnarray}
    while its $s$-Riemann surface ${\bf R}_B /T$ approaches 'smoothly' the topological
     structure of the Riemann surface  corresponding to the first term 
${\bf B}_0(s)$ of the     expansion (\ref{2.1}).

     ${\bf 5}^0$ With respect to its dependence 
on $s$ the field ${\bf B}_0(s)$ satisfies properties ${\bf 1}^0-{\bf 3}^0$ above
     with substitutions $t\to s$ and ${\bf B}(s)\to{\bf B}_0(s)$.
     
Note that condition ${\bf 3}^0$ excludes periodic fields ${\bf B}(t)$ .

     The time-dependent Schr\"odinger equation induced by $H(t)$ takes 
therefore a form
\begin{eqnarray}
\frac{i}{T}\frac{d\Psi(s,T)}{ds}=\frac{1}{2} \mu{\bf B}(sT,T)\cdot{\bf \sigma} 
\Psi(s,T)
\label{2.2}
\end{eqnarray}

     The adiabatic regime of evolution of the wave function $\Psi(s,T)$
 corresponds now to taking a limit $T\to +\infty$ in (\ref{2.2}). 

     The main problem of the adiabatic limit in the considered case is to 
find in this limit the
transition amplitude between the two energy levels of the system for 
$s\to +\infty$ under the
assumptions that $\Psi(-\infty,T)$ coincides with one of the two possible 
eigenstates $\Psi_{\pm}(-\infty,T)$ of $H(-\infty)$
and that there is no level crossing for real $t$ i.e. $\displaystyle\liminf_{-\infty <t <+\infty}B(t)\ge\epsilon>0$.
 Known approximate solutions of this
problem are that of Landau \cite{13} and Zener \cite{14} in a form of
the so 
called Landau-Zener formula and that of Dykhne \cite{15} who have shown that 
such an amplitude should be exponentially small in the limit $T\to +\infty$.
 In the next sections we shall show how
to get an exact (i.e. not approximate) result for this
amplitude as well as its adiabatic limit with the help of the fundamental solutions . 

     A typical way of proceeding when the adiabatic limit is investigated is 
using eigenvectors $\Psi_{\pm}(s,T)$ of $H(sT,T)$ satisfying 
$(\Psi_{\pm},\dot\Psi_{\pm})=0$. Then, such eigenvectors
$\Psi_{\pm}(s,T)$ can be chosen as the following ones
\begin{eqnarray}
\Psi_+(s,T)=e^{-i\int
_0^s\dot\phi\sin^2\frac{\Theta}{2}d\sigma}
\left[\begin{array}{c}\cos\frac{\Theta}{2}\\\sin\frac{\Theta}{2}e^{i\phi}\end{array}\right],\hspace{5mm}
\Psi_-(s,T)=e^{-i\int
_0^s\dot\phi\cos^2\frac{\Theta}{2}d\sigma}
\left[\begin{array}{c}\sin\frac{\Theta}{2}\\-\cos\frac{\Theta}{2}e^{i\phi}\end{array}\right]
\label{2.3}
\end{eqnarray}
where $\Theta$ and $\phi$ are polar and azimuthal angles 
of the vector ${\bf B}(t,T)$, respectively, and dots
over different quantities mean derivatives with respect to $s$-variable.

     The wave function $\Psi(s,T)$ can now be represented as
\begin{eqnarray}
\Psi(s,T)=a_+(s,T)e^{-iT \int_{s'}^sE_+(\xi,T)d\xi}\Psi_+(s,T)
+a_-(s,T)e^{-iT \int_{s'}^sE_-(\xi,T)d\xi}\Psi_-(s,T)
\label{2.4}
\end{eqnarray}
where $s'$ takes $any$ real but fixed value.

The Schr\"odinger equation (\ref{2.2}) can be rewritten in terms of the coefficients $a_{\pm}(s,T)$ as the
following linear system of two equations
\begin{eqnarray}
\dot a_+(s,T)=c(s,T)e^{i\int_{s'}^s\omega(\xi,T)d\xi}a_-(s,T)\nn  \\
\label{2.5}\\
\dot a_-(s,T)=-c^*(s,T)e^{-i\int_{s'}^s\omega(\xi,T)d\xi}a_+(s,T)\nn
\end{eqnarray}
where
\begin{eqnarray}
c(s,T)= - \frac{\dot\Theta}{2} + \frac{i\dot\phi}{2}
\sin\Theta=-\frac{1}{2}\frac{\left[{\bf B}\times\left({\bf
      B}\times{\bf {\dot
        B}}\right)\right]_z}{B^2\sqrt{B_x^2+B_y^2}}+\frac{i}{2}\frac{\left({\bf B}\times{\bf{\dot B}}\right)_z}{ B\sqrt{B_x^2+B_y^2}}\nn\\
\label{2.6}\\
\omega(s,T)=T\left( E_+-E_- \right)-\dot\phi\cos\Theta=\mu
TB-\frac{B_z}{B}\frac{\left({\bf B}\times{\bf{\dot
        B}}\right)_z}{B_x^2+B_y^2}\nn
\end{eqnarray}

     The system (\ref{2.5}) can be rewritten further as the
     following linear system of second order
equations
\begin{eqnarray}
\ddot a_+ -\left(\frac{\dot c}{c}+i\omega\right)\dot a_+ +|c|^2a_+=0
\nn\\
\\
\ddot a_- -\left(\frac{\dot c^*}{c^*}-i\omega\right)\dot a_-
  +|c|^2a_-=0 \nn
\label{2.7}
\end{eqnarray}
where the coefficient functions $a_{\pm}$ decouple from each other
being however still related by (\ref{2.5}).

     By the following transformations 
\begin{eqnarray}
a_+(s,T)=e^{\frac{1}{2}\int_{s'}^s\left(\frac{\dot c}{c}+i\omega
  \right)d\xi}b_+(s,T)\nn\\
\\
a_-(s,T)=e^{\frac{1}{2}\int_{s'}^s\left(\frac{\dot c^*}{c*}-i\omega
  \right)d\xi}b_-(s,T)\nn
\label{2.8}
\end{eqnarray}
we bring the equations (\ref{2.7}) to Schr\"odinger types
\begin{eqnarray}
\ddot b_{\pm}(s,T)+T^2q_{\pm}(s,T)b_{\pm}(s,T)=0
\label{2.9}
\end{eqnarray}
where
\begin{eqnarray}
q_+(s,T)=\frac{1}{T^2}\left[-\frac{1}{4}\left(\frac{\dot
      c}{c}+i\omega\right)^2+|c|^2\right]+\frac{1}{2T^2}\left(\frac{\dot
      c}{c}+i \omega\right)^\cdot\nn\\
\\
q_-(s,T)=\frac{1}{T^2}\left[-\frac{1}{4}\left(\frac{\dot
    c^*}{c*}-i\omega\right)^2+|c|^2\right]+\frac{1}{2T^2}\left(\frac{\dot
      c^*}{c^*}-i \omega\right)^\cdot\nn
\label{2.10}
\end{eqnarray}
so that for real $s$ (and $T$) we have
\begin{eqnarray}
q_-(s,T)=q_+^*(s,T)
\label{2.11}
\end{eqnarray}     

The equations (\ref{2.9}) are now basic for our further analysis since their form is just of the
stationary 1-D Schr\"odinger equation. 

     First let us note that the dependence of the "potential" function
     $q_+(s,T)$ on $T$ is given by
\begin{eqnarray}
q_+(s,T)=\frac{1}{4}\mu^2B^2+\frac{i\mu }{2T}\left[\dot B-B\left(\frac{\dot
  c}{c}-i\dot\phi\cos\Theta\right)\right]+\frac{1}{T^2}\left[-\frac{1}{4}\left(\frac{\dot
  c}{c}-i\dot\phi\cos\Theta\right)^2+|c|^2\right]+\nn\\
\label{2.12}
\\\frac{1}{2T^2}\left(\frac{\dot
      c}{c}-i\dot \phi \cos \Theta\right)^\cdot \hspace{50mm}\nn
\end{eqnarray} 
where the dependence of $B, c, \Theta, \phi$ on $T$ in (\ref{2.12}) is also anticipated. By (\ref{2.12}) we get a
corresponding dependence of $q_-(s,T)$ on $T$. Taking into account (\ref{2.1}) and (\ref{2.6}) it is easy to check
that the last formula provides us with the following type of asymptotic behavior of $q_+(s,T)$
for large $T$:
\begin{eqnarray}
q_+(s,T)=q_+^{(0)}(s)+\frac{1}{T}q_+^{(1)}(s)+\frac{1}{T^2}q_+^{(2)}(s)+\dots
\label{2.13}
\end{eqnarray}

Therefore the above form of dependence of $q_{\pm}(s,T)$ on $T$
permits us to apply to the considered case the method of fundamental
solutions. For this reason we shall start the next section with a
review of basic principles of the method suitably adapted to the
considered case.

\section*{3. Fundamental solutions and their properties}

\hskip+2em Consider first $q_{\pm}(s,T)$ as functions of $s$. They are defined completely by an $s$-dependence
of field ${\bf B}(Ts,T)$. According to our assumptions, the latter is meromorphic on some Riemann
surface ${\bf R}_B/T$. However, by (\ref{2.12}), $q_{\pm}(s,T)$ are
algebraic functions of ${\bf B}$,  ${\bf {\dot B}}$ and ${\bf {\ddot B}}$  and,
therefore, they are also meromorphic functions of $s$ defined again on some other Riemann
surfaces ${\bf R}_{\pm}$ determined by these algebraic dependencies. As it follows from (\ref{2.12}) topological structures
of ${\bf R}_{\pm}$ can be quite complicated. However, in what follows, we are interested in the
adiabatic limit $T \to +\infty$ by which the structure of
${\bf R}_{\pm}$ should be determined for $T$ large
enough basically by the first term $q_+^{(0)}(s)$ of the expansion (\ref{2.13}). In consequence, by (\ref{2.12}), it should be
determined by $\mu{\bf B}^{(0)}(s)$ i.e. by the first term of the
expansion (\ref{2.1}). The structure of ${\bf R}_{\pm}$ can turn out
to be much simpler in this limit.
     Despite this supposed complexity of $q_{\pm}(s,T)$  and of their Riemann surfaces we shall
introduce and discuss the fundamental solutions to the equations (\ref{2.9}) without simplifications.
We shall do it for the $q_+(s,T)$ case of (\ref{2.12}). An extension of the
discussion to the $q_-(s,T)$ case will
be obvious.

     A standard way of introducing the fundamental solutions is a construction of a Stokes graph
\cite{8,9,10} related to a given $q_+(s,T)$. Such a construction, according to Fr\"oman and Fr\"oman \cite{8}
and Fedoriuk \cite{9}, can be performed in the following way \cite{10}. 

     Let $Z$ denote a set of all the points of ${\bf R}_+$ at which $q_+(s,T)$ has its single or double poles.
Let $\delta(x)$ be a meromorphic function on ${\bf R}_+$, the unique singularities of which are double poles at
the points collected by $Z$ with coefficients at all the poles equal to $1/4$ each. (In a case when ${\bf R}_+$ is simply a complex plain the latter
function can be constructed in general with the
help of the Mittag-Leffler theorem \cite{17}. But for a case of
branched ${\bf R}_+$ the general procedure
is unknown to us).
     Consider now a function
\begin{eqnarray}
\tilde{q}_+(s,T)=q_+(s,T)+\frac{1}{T^2}\delta(s)
\label{3.1}
\end{eqnarray}

     The presence and the role of the $\delta$-term in (\ref{3.1}) are explained below. This term contributes
to (\ref{3.1}) if and only when the corresponding 'potential' function $q_+(s,T)$ contains simple or
second order poles. (Otherwise the corresponding $\delta$-term is put to zero). It is called the Langer
term \cite{10,18}.

     The Stokes graph corresponding to the function $\tilde{q}_+(s,T)$ consists now of Stokes lines
emerging from roots (turning points) of $\tilde{q}_+(s,T)$. Stokes lines satisfy one of the following
equations: 
\begin{eqnarray}
\Im \int_{s_i}^s\sqrt{\tilde{q}_+(\xi,T)}d \xi=0
\label{3.2}
\end{eqnarray}
with $s_i$ being a root of $\tilde{q}_+(s,T)$. We shall assume further a generic situation when all the roots
$s_i$ are simple.

     Stokes lines which are not closed end at these points of ${\bf R}_+$ (i.e. have the latter points
as their boundaries) for which the action integral in (\ref{3.2}) becomes infinite. Of course such points
are singular for $\tilde{q}_+(s,T)$ and they can be its finite poles or its poles lying at an infinity. 

     Each such a singularity $z_k$ of $\tilde{q}_+(s,T)$ defines a domain called a sector. This is the
connected domain of  ${\bf R}_+$ bounded by Stokes lines and $z_k$
itself. The latter is also
a boundary for the Stokes lines or being an isolated boundary point of the sector (as it is in the
case of the second order pole). 

     In each sector the LHS in (\ref{3.2}) is only positive or only negative. 

     Consider now equation (\ref{2.9}) for $b_+(s,T)$. Following Fr\"oman and Fr\"oman in each sector $S_k$ having a singular point $z_k$ at its
boundary one can
define a solution of the form:
\begin{eqnarray}
b_{+,k}(s,T) = \tilde{q}_+^{-\frac{1}{4}}(s,T){\cdot}
e^{\sigma i T W(s,T)}\chi_{+,k}(s,T) &
& k = 1,2,\ldots
\label{3.3}
\end{eqnarray}
where
\begin{eqnarray}
\chi_{+,k}(s,T) = 1 + \sum_{n{\geq}1}
\left( -\frac{\sigma}{2iT} \right)^{n} \int_{z_k}^{s}d{\xi_{1}}
\int_{z_k}^{\xi_{1}}d{\xi_{2}} \ldots 
\int_{z_k}^{\xi_{n-1}}d{\xi_{n}}
\Omega(\xi_{1})\Omega(\xi_{2}) \ldots \Omega(\xi_{n})\times\nn\\
\\ 
\left( 1 - e^{-2\sigma iT{(W(s)-W(\xi_{1}))}} \right)
\left(1 - e^{-2\sigma iT{(W(\xi_{1})-W(\xi_{2}))}} \right)
\cdots
\left(1 - e^{-2\sigma iT{(W(\xi_{n-1})-W(\xi_{n}))}}
\right)\nn
\label{3.4}
\end{eqnarray}
with
\begin{eqnarray}
\Omega(s,T) = \frac{\delta(s)}{\tilde{q}_+^{\frac{1}{2}}(s,T)} - 
{\frac{1}{4}}{\frac{\tilde{q}_+^{\prime\prime}(s,T)}
{\tilde{q}_+^{\frac{3}{2}}(s,T)}} +
{\frac{5}{16}}{\frac{\tilde{q}_+^{\prime 2}(s,T)}
{\tilde{q}_+^{\frac{5}{2}}(s,T)}}
\label{3.5}
\end{eqnarray}
and
\begin{eqnarray}
W(s,T) = \int_{s_{i}}^{s} \sqrt{\tilde{q}(\xi,T)}d\xi
\label{3.6}
\end{eqnarray}
where $s_i$ is a root of $\tilde{q}(s,T)$  lying at the boundary of $S_k$.

     In (\ref{3.3}) and (\ref{3.4}) a sign of $\sigma$ (=$\pm 1$) and an integration path are chosen in such a way
to have:
\begin{eqnarray}
\sigma \Im \left(W(\xi_{j}) - W(\xi_{j+1}) \right) \leq 0
\label{3.7}
\end{eqnarray} 
for any ordered pair of integration variables (with $\xi_0=s$). Such
an integration path is then
called canonical. Of course, the condition (\ref{3.7}) means that $b_{+,k}(s,T)$ vanishes in its sector when
$s\to z_k$ along the canonical path.
     The Langer $\delta$-term appearing in (\ref{3.1}) and (\ref{3.5}) is necessary to ensure all the integrals
in (\ref{3.4}) to converge when $z_k$ is a first or a second order pole of  $\tilde{q}(s,T)$ or when the solutions
(\ref{3.3}) are to be continued to such poles. As it follows from (\ref{3.5}) each such pole $z_k$ demands
a contribution to $\delta(s)$ of the form $\left(2(s-z_k)\right)^{-2}$, what has been already assumed in the corresponding
construction of $\delta(s)$.

\section*{4. The adiabatic limit in the fundamental solution approach}
\hskip+2em     Consider now the consequences of taking the large-$T$ limit for the above description. We
assume that for a given $\tilde{q}_+(s,T)$ and its Riemann surface  ${\bf R}_+$ the corresponding Stokes graph ${\bf G}_+$
is drawn. It is drawn, of course, on the Riemann surface $\sqrt{{\bf R}_+}$ corresponding to $\sqrt{\tilde{q}_+(s,T)}$. 

     First let us notice that singular points of $\tilde{q}_+(s,T)$ such as its branch points and poles
depend in general on $T$. For both kinds of these singularities this
also means a dependence
on $T$ of jumps of $\tilde{q}_+(s,T)$ on its cuts as well as the $T$-dependence of coefficients of its poles.

     According to the property ${\bf 4}^0$ of the magnetic field ${\bf B}$ (see Sec. 2) we can expect that the
singular structure of $\tilde{q}_+(s,T)$, i.e. positions of its roots and poles, as well as the cut jumps and
pole coefficients, change smoothly in this limit to their final positions and values respectively.
This limit structure is defined by the singularity structure of $\tilde{q}_+^{(0)}(s,T)$ (see expansion
(\ref{2.13})). Therefore, both the topology of $\sqrt{{\bf R}_+}$ and the associated Stokes graph ${\bf G}_+$ change
accordingly to coincide eventually with the Riemann surface
$\sqrt{{\bf R}_+^{(0)}}$ and with the Stokes graph
${\bf G}_+^{(0)}$ corresponding to $\sqrt{\tilde{q}_+^{(0)}(s,T)}$ . This limit structure can be achieved in the following
ways:

     {\bf a}. some of branch points and poles of
     $\tilde{q}_+(s,T)$ escape to infinities of ${\bf R}_+$;

      {\bf b}. some of branch points and poles of $\tilde{q}_+(s,T)$ approach the respective singularities
     of $\tilde{q}_+^{(0)}(s,T)$;

     {\bf c}. some of branch points and poles of $\tilde{q}_+(s,T)$ disappear because their respective
     jumps and coefficients vanish in the limit $T\to  +\infty$.

     Being more specific we expect that for $T$ large enough a set ${\bf S}_+$ of all singular
points of $\tilde{q}_+(s,T)$ (i.e. containing all its branch points and poles) consists of three well
separated subsets ${\bf S}_+^{inf}$, ${\bf S}_+^{van}$ and ${\bf
  S}_+^{fin}$. The points of ${\bf S}_+^{inf}$ run to infinities of
${\bf R}_+$ when $T\to +\infty$.
Those of ${\bf S}_+^{van}$ disappear in this limit while those of ${\bf
  S}_+^{fin}$ coincide in this limit with the set ${\bf S}_+^{(0)}$
of the singular points of $\tilde{q}_+^{(0)}(s,T)$ .

     Let us remove the
     points contained in ${\bf S}_+^{inf}\cup{\bf S}_+^{van}$ from the Riemann surface ${\bf R}_+$ , 
i.e. let us consider these points as regular for
     $\tilde{q}_+(s,T)$. Then ${\bf R}_+$ will transform
     into ${\bf R}_+^{fin}$ - a
Riemann surface which singular points coincide with those of the set ${\bf
  S}_+^{fin}$. 

     Together with the previous operation let us remove from $\sqrt{{\bf R}_+}$ also the Stokes lines
generated by the points of ${\bf S}_+^{inf}\cup{\bf S}_+^{van}$ so that the remaining Stokes lines can be uniquely continued
to form the Stokes graph ${\bf G}_+^{fin}$ generated by the set ${\bf S}_+^{fin}$. It is clear that the graph ${\bf G}_+^{fin}$ coincides
with ${\bf G}_+^{(0)}$ in the limit $T\to +\infty$. 

     The above two operations will be called the adiabatic limit reduction or simply the
reduction operation. 

     As we have mentioned earlier there is a set of sectors and a corresponding set of
fundamental solutions defined in them associated with the graph  ${\bf G}_+$. By the reduction operation
both sets can be reduced i.e. under this operation some sectors of
${\bf G}_+$ transform into corresponding sectors of ${\bf G}_+^{fin}$ whereas the others disappear. Obviously, the latter sectors are those
which disappear when the limit $T\to +\infty$ is taken.

     The following assumption should stabilize the corresponding results obtained with the
help of the fundamental solution method.

     ${\bf 6}^0$ Among a full set of fundamental solutions associated with the Stokes graph ${\bf G}_+$ there
     is a subset of them associated with graph ${\bf G}_+^{fin}$ which allows us to solve the basic
     problem of the adiabatic transition and which is invariant under the reduction operation.

     The dynamical systems described by the Hamiltonian $H(t)$
     satisfying assumption ${\bf 6}^0$
will be called the adiabatic limit reducible (ALR-)systems.

     The above assumption means that to solve the problem of the adiabatic transitions in the
ALR-system we can first perform the reduction operation and next work with the simplified
Stokes graphs ${\bf G}_+^{fin}$. A set of fundamental solutions associated with this graph which can be used
to solve the problem considered coincide with the corresponding ones of the full graph ${\bf G}_+$. The
procedure used to construct a solution of the problem with the help of the latter graph is not
affected by the reduction operation, i.e. it looks the same when the simplified graph ${\bf G}_+^{fin}$ is used
instead of ${\bf G}_+$. Therefore the aim of the reduction operation is to make easier choosing the
proper set of fundamental solution solving the problem. The results obtained in this way can
be still exact if the integration paths taken on the graph ${\bf G}_+^{fin}$ can be mapped properly on the
Stokes graph ${\bf G}_+$ restoring in this way the exact condition of the problem. However, if such a
map is not known or is difficult to construct (because of the complicated structure of graph
${\bf G}_+$) the result obtained in this way can be considered only as an approximation i.e. valid only
in the limit $T\to+\infty$. 

     According to the above assumptions we can conclude from (\ref{2.12}) and (\ref{2.13}) that there
is one-to-one correspondence between the Stokes graphs ${\bf G}_+$ and ${\bf G}_+^{(0)}$ and the corresponding sets
${\bf S}_+^{fin}$ and ${\bf S}_+^{(0)}$. Namely, this correspondence is built by aggregations (blobs) of singular points of
${\bf S}_+^{fin}$, i.e. the branch points and poles of $\tilde{q}_+(s,T)$, which are transformed into single points of
${\bf S}_+^{(0)}$ when the limit $T\to+\infty$ is taken. Also there are sheaves of Stokes lines of ${\bf G}_+^{fin}$ emerging
from the blobs and transformed into single lines of ${\bf G}_+^{(0)}$ in the same limit.

     Therefore in the limit $T\to+\infty$ we can eventually consider for 
potentials (\ref{2.12}) Stokes graphs corresponding to first terms 
$q_{\pm}^{(0)}(s)$ of the asymptotic expansions for $q_{\pm}(s,T)$.
The first terms of the asymptotic expansions corresponding to $q_{\pm}^{(0)}(s)$ 
and $q_{\pm}(s,T)$ are the same in this limit and equal, according to (\ref{2.1}), 
to $\frac{1}{4}\mu^2{\bf B}_0^2(s)$. 

     Let us note that properties ${\bf 1}^0-{\bf 6}^0$ above can be satisfied
     by the field ${\bf B}$ for which ${\bf B}^2$ 
is a $meromorphic$ function of $t$. We shall assume just such a
     dependence of ${\bf B}$ on $t$ and of the
corresponding rescaled field ${\bf B}(sT,T)$ on $s$. However, for simplicity, instead of
continuing our considerations in their most general form we shall investigate first a particular
example of the field ${\bf B}(t,T)$ which, as it seems to us, will illustrate our method in a satisfactory
way. 

\section*{5. The Nikitin model of the atom-atom scattering}

\hskip+2em     The model of Nikitin \cite{12} describes the scattering
A*+B$\rightarrow$A+B+$\Delta\epsilon$ of the exited atom
A* moving with a small velocity $v$ with the impact parameter $b'$ and scattered by the atom
B. The interaction between the atoms is of the dipol-dipol type. The latter example was
analyzed in the context of the adiabatic limit $v\to0$ also by Joye \underline
  {et al} \cite{4}.

The Hamiltonian for this system reads (\cite{11}, paragraph 9.3.2 and \cite{12}):
\begin{eqnarray}
H(R)=\left[\begin{array}{cc}\frac{\Delta\epsilon}{2}&\frac{C}{R^3}\\\frac{C}{R^3}&-\frac{\Delta\epsilon}{2}\end{array}\right]
\label{5.1}
\end{eqnarray}
where $\Delta\epsilon$ and $C$ are constants and $R=\sqrt{{b'}^2+v^2t^2}$ is the distance between the atoms. Introducing
$d=(2C/\Delta\epsilon)^{\frac{1}{3}}$ as a natural distant unit for this case and $T=d/v$ as the corresponding adiabatic
parameter and rescaling: $t\to sT$ and $b'\to bd$ we get from (\ref{5.1}):
\begin{eqnarray}
H(s)=\frac{\Delta\epsilon}{2}\left[\begin{array}{cc}1&\frac{1}{(b^2+s^2)^{\frac{2}{3}}}\\\frac{1}{(b^2+s^2)^{\frac{2}{3}}}&-1\end{array}\right]
\label{5.2}
\end{eqnarray}

     In the 'magnetic field' language we have of course ${\bf
       B}(sT,T)=\left(\left(b^2+s^2\right)^{-\frac{3}{2}},0,1\right)\frac{\Delta\epsilon}{\mu}$ so that
all the assumptions ${\bf 1}^0-{\bf 6}^0$ above are satisfied with ${\bf
       B}^{\pm}(T)={\bf B}^{\pm}(\pm
       \infty,T)=(0,0,1)\frac{\Delta\epsilon}{\mu}$. Since in the
considered case $\phi(s)\equiv 0$ then for the corresponding quantities defined by (\ref{2.6}) and (\ref{2.12}) we
get:
\begin{eqnarray}
c=\frac{3}{2}\frac{s\left(b^2+s^2\right)^{\frac{1}{2}}}{1+\left(b^2+s^2\right)^3},\hspace{10mm}\omega=T\Delta\epsilon
\left(1+\frac{1}{\left(b^2+s^2\right)^3}\right)^{\frac{1}{2}}\hspace{20mm}\nn\\
\nn\\
q_{\pm}(s,T)=\left[\frac{\Delta\epsilon}{2}\left(1+\frac{1}{\left(b^2+s^2\right)^3}\right)^{\frac{1}{2}}\pm\frac{i}{2T}\left(\frac{6s(b^2+s^2)^2}{1+(b^2+s^2)^3}-\frac{s}{b^2+s^2}-\frac{1}{s}\right)\right]^2
-\nn\\
\label{5.3}\\
\frac{3}{2}\frac{i\Delta\epsilon}{T}\frac{s}{\left(1+(b^2+s^2)^3\right)^\frac{1}{2}(b^2+s^2)^\frac{5}{2}}-\hspace{40mm}\nn\\
\nn\\
\frac{1}{2T^2}\left[\frac{2s^2+b^2(b^2+s^2)}{s^2(b^2+s^2)}-\frac{3}{2}\;\frac{\;4(b^2+s^2)^4(s^2-b^2)-4(b^2+s^2)(b^2+5s^2)+3s^2(b^2+s^2)}{\left(1+(b^2+s^2)^3\right)^2}\right]\nn
\end{eqnarray}

     Equations (\ref{5.3}) show that in the limit $T\to + \infty$ the Stokes graph for the considered
problem is determined by the function
\begin{eqnarray}
q^{(0)}(s,T)=\frac{(\Delta\epsilon)^2}{4}\left(1+\frac{1}{\left(b^2+s^2\right)^3}\right)
\label{5.4}
\end{eqnarray}
The graph is shown on Fig.1.

  Each $q_{\pm}(s,T)$ has 40 roots, five branch points at
  $s=\pm ib$ and at $s=s_k=\pm \left(e^{\frac{(2k+1)\pi i}{3}}-b^2\right)^{\frac{1}{2}}$
, $k=1,2,3$, as well as two poles at $s=0$. Therefore only six roots of
  $q^{(0)}(s,T)$ at $s=s_k$, $k=1,2,3$ and its two poles at $s=\pm ib$ look encouraging. Nevertheless, we shall consider first the
case without any approximations.

     At first glance the Stokes graphs corresponding to the functions
     $q_{\pm}(s,T)$ seem to be quite complicated. However it can be handled in the following way. 

     Functions $q_{\pm}(s,T)$ are determined on two sheeted
     Riemann surfaces ${\bf R}_{\pm}$ respectively
with the branch points at $s=\pm ib$ and at $s=s_k$, $k=1,2,3$ and with $40$ roots distributed into halves on each
sheet of the surfaces. Therefore the Riemann surfaces $\sqrt
     {\bf R}_{\pm}$ corresponding to $\sqrt{q_\pm(s,T)}$
 (it will turn out that it is not necessary to introduce to the latter functions the
corresponding Langer terms) are four-sheeted with these $40$ roots being square root branch
points on them. When $T\to +\infty$ only six of these branch points survive coinciding with
the six roots of $q^{(0)}(s,T)$ at $s=\pm s_k$, k=1,2,3 whereas ${\bf R}_{\pm}$ transforms into the complex $s$-plane since
the branch points of $q_{\pm}(s,T)$ at $s=\pm ib$ disappear, being transformed into the second order poles
of $q^{(0)}(s,T)$. It is easy to check however that for finite but large $T$ these six roots of $q^{(0)}(s,T)$ are
each split initially into two as. The split is the result of the square root branch
points at $s=\pm ib$ to which the
recovering of the finite $T$ transforms the poles of $q^{(0)}(s,T)$ at the same points. The two copies
of each of these six roots lie of course on different sheets of ${\bf R}_{\pm}$. Next, each of these 12 roots
is still split into three by the same reason of finiteness of $T$. In this way, on each of
the two sheets of ${\bf R}_{\pm}$ there are $36$ roots grouped by
     three around their limit $s=\pm s_k$,
$k=1,2,3$ achieved for $T\to +\infty$.

     The remaining four roots of $q_{\pm}(s,T)$ are displaced in two pairs, one pair on each sheet
of ${\bf R}_{\pm}$, close to the points $s=0$ at which the second order poles of $q_{\pm}(s,T)$ are localized. When
$T\to +\infty$ the roots in each pair collapse into $s=0$ multiplying the corresponding second order
poles and thus causing mutual cancellations of the latter and
themselves in this limit.

     Now we shall focus our attention on
     the Stokes graph $\bf G _-$
generated by $q_-(s,T)$ on the first sheet of ${\bf R}_-$ as well as on the remaining
ones. It looks as in Fig.2. 
The Stokes graph $\bf G _+$ corresponding to $q_+(s,T)$ can be obtained from $\bf G _-$ by complex
conjugation of the latter. On the figure the wavy lines denote the cuts corresponding to the branch points of the fundamental solutions
defined on ${\bf R}_-$. The sheet on Fig.2 cut along the wavy lines defines a domain where all the
fundamental solutions $b_{-,1}(s,T),...,b_{-,\bar 2}(s,T)$ defined in
the corresponding sectors $S_1,...,S_{\bar 2}$ (shown
in the figure) are holomorphic.

\vskip 12pt
\begin{tabular}{cc}
\psfig{figure=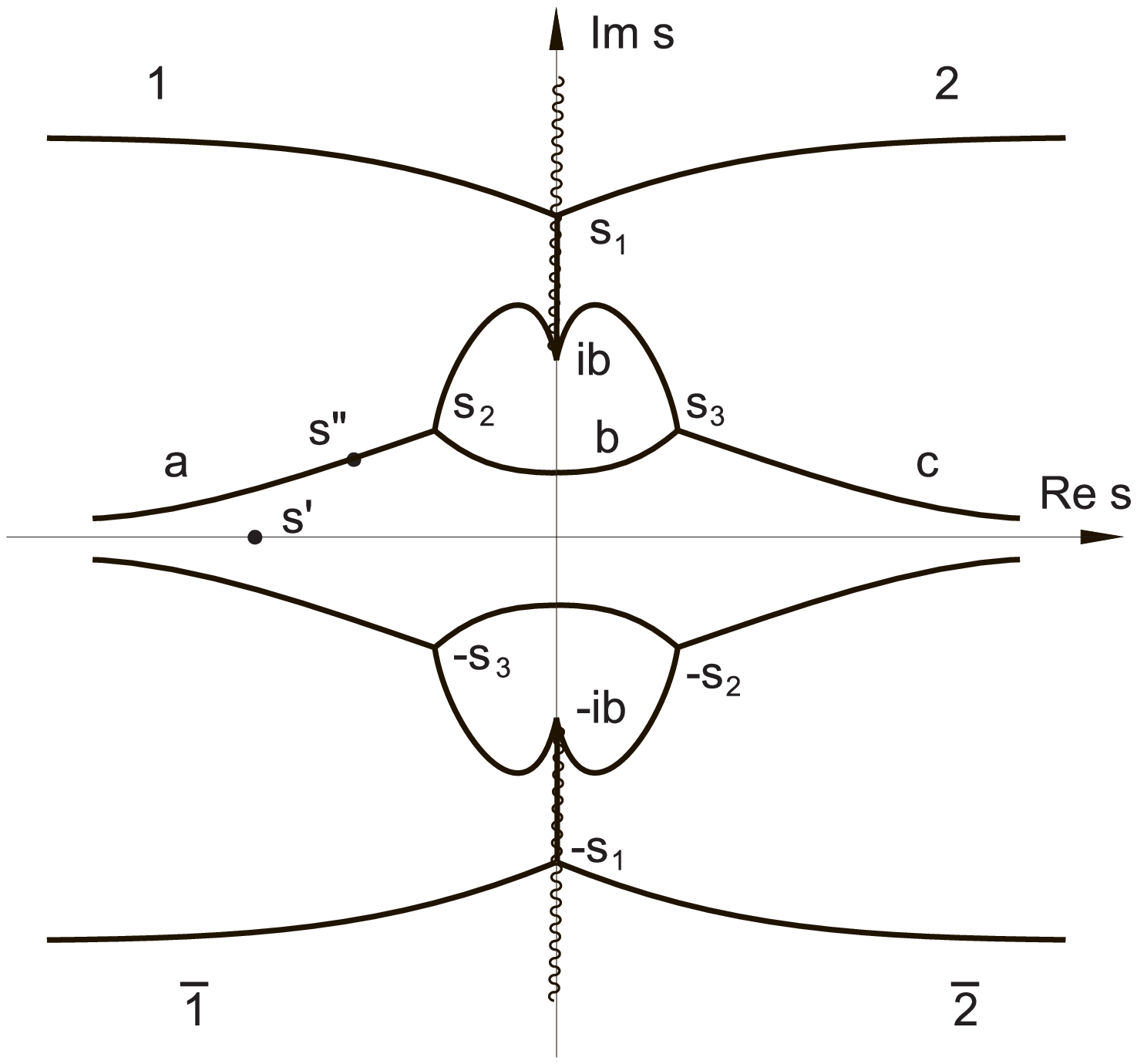,width=7cm}&\psfig{figure=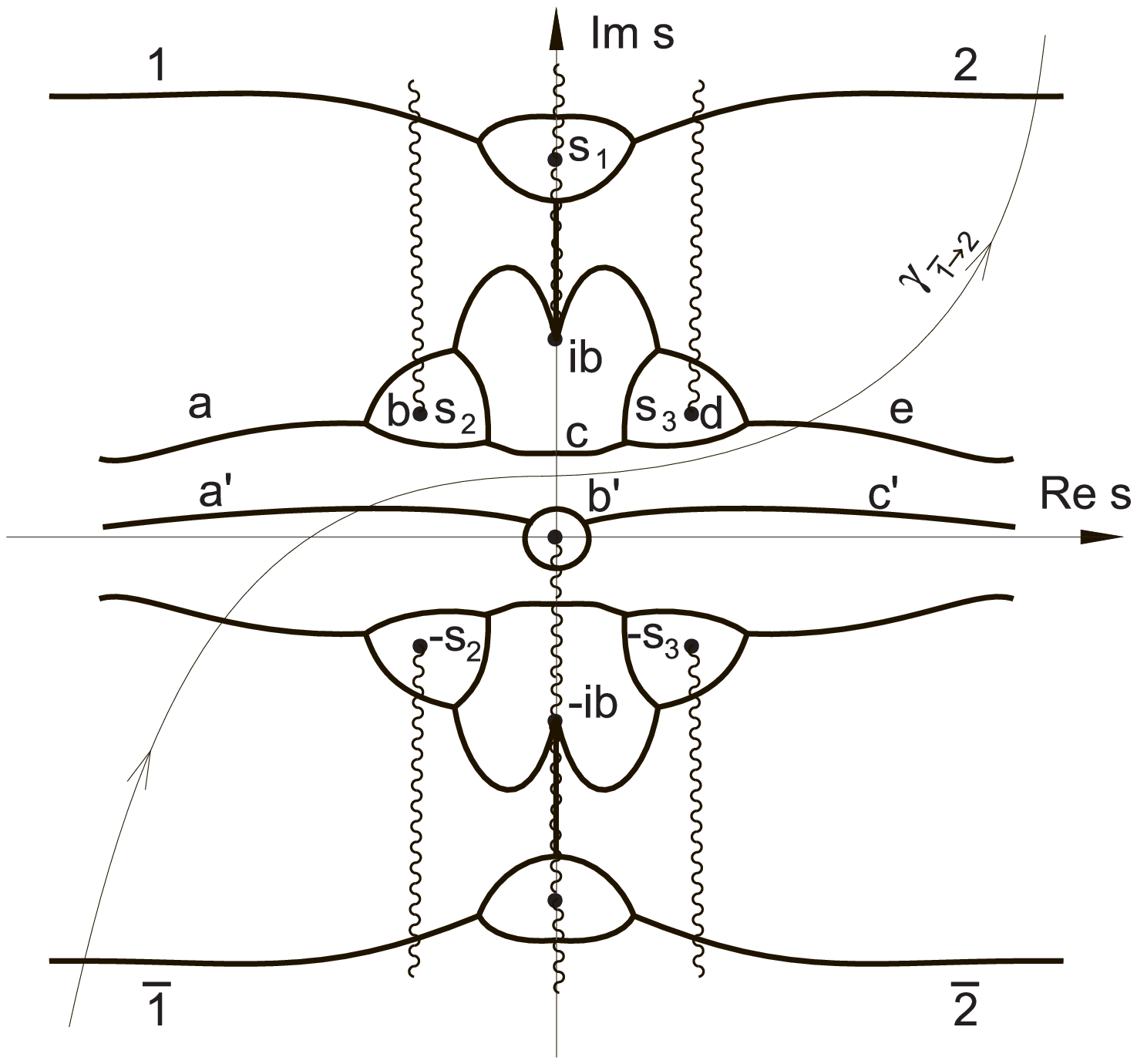,width=7cm} \\
Fig.1 The Stokes graph corresponding& Fig.2
The Stokes graph corresponding\\
      to 'potential' (\ref{5.4})&       to 'potential' $q_-(s,T)$ of
(\ref{5.3})  
\end{tabular}
\vskip 12pt
 
     According to our earlier description of the behavior of the Riemann surface$\sqrt{R_+}$  when
$T\to +\infty$ the set ${\bf S} _-^{inf}$ corresponding to the considered case is empty, ${\bf S} _-^{van}$ contains four points at
$s=0$ on each of the four sheets of $\sqrt{{\bf R}_-}$ (these four points correspond to the second order poles
of $q_-(s,T)$) and the four branch points close to $s=0$, while ${\bf S}_-^f$ contains all the remaining singular
points of $\sqrt{q_-(s,T)}$.
     Now, for our case, the solution of the problem stated in this paper is simple.
Namely, it can be found in the following steps:

       {\bf i}. take a linear combination of the fundamental solutions
     $b_{-,1}(s,T)$ and $b_{-,\bar 1}(s,T)$ to
     construct the amplitude $a_-(s,T)$ with the desired property at
     $s=-\infty$, i.e. $\displaystyle\lim_{s\to -\infty}|a_-(s,T)|=0$. This
     amplitude is defined in this way up to a multiplicative constant;

      {\bf ii}. use the equation (\ref{2.5}) to construct $a_+(s,T)$ and adjust the constant mentioned earlier
     so that the limit $\displaystyle\lim_{s\to -\infty}|a_+(s,T)|=1$ can be satisfied;

     {\bf iii}. continue canonically $a_-(s,T)$ along the real s-axis
      with the help of the solutions $b_{-,1}(s,T)$ and $b_{-,\bar 1}(s,T)$      using to this goal the remaining fundamental solutions if necessary;

      {\bf iv}. calculate the limit $s\to +\infty$; and

      {\bf v}. calculate the adiabatic limit $T\to +\infty$.

     According to (\ref{2.8}) and to the first of the above steps we have:
\begin{eqnarray}
a_-(s,T)=Aq_-^{-\frac{1}{4}}(s,T)e^{\int_{s'}^s\frac{1}{2}\left(\frac{\dot
      c}{c}-i\omega\right)(\sigma,T)d\sigma+iT\int_{s_0}^sq_-^\frac{1}{2}(\sigma,T)d\sigma}\chi_{\bar
      1}(s,T)\nn\\
\\
+Bq_-^{-\frac{1}{4}}(s,T)e^{\int_{s'}^{s}\frac{1}{2}\left(\frac{\dot
      c}{c}-i\omega\right)(\sigma,T)d\sigma-iT\int_{s_0}^s q_-^\frac{1}{2}(\sigma,T)d\sigma}\chi_1(s,T)\nn
\label{5.5}
\end{eqnarray}
where $s'$ is any point on the real axis which is regular for the integrand whilst $s_0$ is the one from
the infinite strip bounded by the Stokes line $abcde$ from one side and by $a'b'c'$ from the other
(see Fig. 2), being also an arbitrary but regular point for all the integrands. The choice of
signatures in (\ref{5.5}) was done due to the fact that $\Re\left(iT\int_{s_0}^{s}q_-^{\frac{1}{2}}d\sigma\right)$  is positive (for $s$ sufficiently
large) for the sector $S_1$ and negative for $S_{\bar 1}$ . The latter property follows from the fact that
according to (\ref{5.5}) and the Stokes graph on Fig. 2 we have on the first
sheet of $\sqrt{{\bf R}_-}$:
sgn$\left(\Re\sqrt{q_-^{\frac{1}{2}}(s,T)}\right)$=sgn$(s)$ for
$s\to\pm\infty$ along the real axis.

      If further we take into account the following asymptotic behavior of the relevant quantities on the real axis:
\begin{eqnarray}
\frac{1}{2}\left[\frac{\dot
    c}{c}-i\omega\right]+iT\sqrt{q_-}\sim-iT\Delta\epsilon-\frac{4}{s},
    s\to
    -\infty\nn\\
\\
\frac{1}{2}\left[\frac{\dot c}{c}-i\omega\right]-iT\sqrt{q_-}\sim O(\frac{1}{s^8}),\;\;\;\;\;\;\;\;s\to-\infty\nn
\label{5.6}
\end{eqnarray}
then we can conclude that $B=0$ in (\ref{5.5}).

     To fix the value of the constant A in (\ref{5.5}) we can use the second of relations (\ref{2.5})
and apply the condition mentioned in the second step of the procedure
     i.e. $\displaystyle\lim_{s\to-\infty}\left(-\frac{1}{c(s,T)}\cdot\right.$
$\left.e^{i\int_{s'}^s\omega
     d\sigma}\dot a_-(s,T)\right)=1$
 to get:
\begin{eqnarray}
A=\frac{1}{T\Delta\epsilon}\sqrt{\frac{\Delta\epsilon}{2}}\exp\left(-\int_{s'}^{s_0}i\omega
        ds+\int_{-\infty}^{s_0}\left[-\frac{1}{2}\left(\frac{\dot
        c}{c}-i\omega\right)+iT\sqrt{q_-}\right]ds+\ln c(s_0)\right)
\label{5.7}
\end{eqnarray}

     Therefore, for the amplitude $a_-(s,T)$ we obtain finally:
\begin{eqnarray}
a_-(s,T)=\frac{1}{T\Delta\epsilon}\sqrt{\frac{\Delta\epsilon}{2}}q_-^{-\frac{1}{4}}(s,T)\exp\left(-\int_{s'}^{s_0}i\omega
  ds+\int_{-\infty}^{s_0}\left[-\frac{1}{2}\left(\frac{\dot
        c}{c}-i\omega\right)+iT\sqrt{q_-}\right]ds+\right.\nn\\
\\
\left.\ln{c(s_0)}+\int_{s_0}^{s}\left[\frac{1}{2}\left(\frac{\dot
  c}{c}-i\omega\right)+iT\sqrt{q_-}\right]d\sigma\right)\chi_{\bar 1}(s,T)\;\;\;\;\;\;\;\;\nn
\label{5.8}
\end{eqnarray}

     Now we can take the limit $s\to+\infty$ in the above formula, continuing  along the
canonical path $\gamma _{\bar 1\to 2}$ shown in Fig.2, to get:
\begin{eqnarray}
a_-(+\infty,T)=\frac{1}{iT\Delta\epsilon}\exp\left(-\int_{s'}^{s_0}i\omega
        ds+\int_{-\infty}^{s_0}\left[-\frac{1}{2}\left(\frac{\dot
        c}{c}-i\omega\right)+iT\sqrt{q_-}\right]ds+\right.\nn
\\
\\\left.\ln{c(s_0)}+\int_{s_0}^{+\infty}\left[\frac{1}{2}\left(\frac{\dot
        c}{c}-i\omega\right)+iT\sqrt{q_-}\right]d\sigma\right)\chi_{\bar
        1\to 2}(T)\;\;\;\;\;\;\;\nn
\label{5.9}
\end{eqnarray}

     The apparent $s_0$-dependence in the above formula is illusive. We can use this fact to
calculate the integrals in the exponent most accurately. First let us note that we cannot disjoint
totally the integrations in the two infinite integrals since the diverging contributions of the three
terms in both of these integrals cancel mutually at the corresponding infinities, making the
integrals convergent. We can however take as the integration paths for these two integrals the Stokes lines $abc$ on Fig.
1 and $abcde$ on Fig. 2. Namely, let the points $s_L$ on line $a$
and $s_R$ on line $e$ be arbitrarily close to the corresponding infinities of the real axis. Let
further points $s_L'$ and $s_R'$ be the points on the Stokes lines $a$ and $c$ of Fig. 1 respectively.
We choose the latter points to lie on the anti Stokes lines of Fig. 1 which pass by the respective
points $s_L$ and $s_R$. Then the integral in the exponential of formula (\ref{5.9}) can be rewritten as:
\begin{eqnarray}
I\equiv-\int_{s'}^{s_0}i\omega
ds+\int_{-\infty}^{s_0}\left[-\frac{1}{2}\left(\frac{\dot c}{c}-i\omega\right)+iT\sqrt{q_-}\right]ds+\ln{c(s_0)}\nn
\\\nn
\\
+\int_{s_0}^{+\infty}\left[\frac{1}{2}\left(\frac{\dot
    c}{c}-i\omega\right)+iT\sqrt{q_-}\right]ds=-\int_{s'}^{s''}i\omega ds\nn
\\\nn
\\
+\int_{-\infty}^{s_L}\left[-\frac{1}{2}\left(\frac{\dot
      c}{c}-i\omega\right)+iT\sqrt{q_-}\right]ds +
      \frac{1}{2}\int_{s_L}^{s'_L}i\omega ds+\frac{1}{2}\ln{c(s_L)}
\label{5.10}
\\\nn
\\
+\int_{s_R}^{+\infty}\left[\frac{1}{2}\left(\frac{\dot
      c}{c}-i\omega\right)+iT\sqrt{q_-}\right]+\frac{1}{2}\int_{s_R}^{s'_R}i\omega
      ds+\frac{1}{2}\ln{c(s_R)}\nn
\\\nn
\\
+\int_{s_L}^{s_R}iT\sqrt{q_-}ds+\frac{1}{2}\int_{s'_L}^{s''}i\omega ds-\int_{s''}^{s'_R}i\omega ds\nn
\end{eqnarray}
where the last three integrals run along the respective Stokes lines and, therefore, are purely
imaginary. Point $s''$ in the above formula is an arbitrary point of the Stokes line $abc$ on
Fig. 1.

     We are interested mainly in the transition probability defined by
     amplitude $a_-(+\infty,T)$ for which only the real part of
     the integral ${\bf I}$ is important. Formula (\ref{5.10}) gives for it:
\begin{eqnarray}
\Re {\bf I}=-\Re\int_{s'}^{s''}i\omega
ds+\Re\int_{-\infty}^{s_L}\left[-\frac{1}{2}\left(\frac{\dot
      c}{c}-i\omega\right)+iT\sqrt{q_-}\right]ds+\frac{1}{2}\int_{s_L}^{s'_L}i\omega ds+\frac{1}{2}\Re\ln{c(s_L)}\nn
\\
\\
+\Re\int_{s_R}^{+\infty}\left[\frac{1}{2}\left(\frac{\dot
      c}{c}-i\omega\right)+iT\sqrt{q_-}\right]ds+\frac{1}{2}\int_{s_R}^{s'_R}i\omega
      ds+\frac{1}{2}\Re\ln{c(s_R)}\nn
\label{5.11}
\end{eqnarray}

     We can now calculate $\Re{\bf I}$ taking in (\ref{5.11}) the limits
     $s_L\to-\infty$ and $s_R\to+\infty$ along the corresponding Stokes lines. We get in this way:
\begin{eqnarray}
\Re {\bf I}=-\Re\int_{s'}^{s''}i\omega
ds+\frac{1}{2}\lim_{s_L\to-\infty}\left(\int_{s_L}^{s'_L}i\omega
  ds+\Re\ln{c(s_L)}\right)+\nn
\\
\label{5.12}
\\
\frac{1}{2}\lim_{s_R\to+\infty}\left(\int_{s_R}^{s'_R}i\omega ds+\Re\ln{c(s_R)}\right)=-\Re\int_{s'}^{s''}i\omega ds+\ln{\frac{3}{2}}\nn
\end{eqnarray}

     The limits in (\ref{5.12}) can be obtained by estimating the asymptotic behaviour of the
differences $s'_{L,R}-s_{L,R}$ and the corresponding functions when $|s|\to\infty$  along the Stokes lines, for
which direct calculation gives:
\begin{eqnarray}
s'_{L,R}-s_{L,R}\sim-\frac{4i}{T\Delta\epsilon}\ln{|s|}-\frac{5b^2i}{2T\Delta\epsilon|s|^2}-\frac{i\ln{a_{L,R}}}{T\Delta\epsilon}\nn
\\\nn
\\
i\omega\sim iT\Delta\epsilon\left(1+\frac{1}{2s^6}\right)
\label{5.13}
\\\nn
\\
\Re\ln{c(s)}\sim\ln{\frac{2}{3}}-4\ln{|s|}-\frac{5b^2}{2|s|^2}\nn
\end{eqnarray}
where constants $a_{L,R}$ are also independent of T and can be estimated exactly only when the
exact equations of the Stokes lines $abc$ of Fig.1 and $abcde$ of Fig.2 are
known.

     The imaginary part of the integral ${\bf I}$ can be calculated as the following limit:
\begin{eqnarray}
\gamma(T)\equiv\Im{\bf I}=\lim_{s_{L,R}\to\mp\infty}\Im\left(\frac{1}{2}\ln{c(s_L)}+\frac{1}{2}\ln{c(s_R)}\right.\;\;\;\;\;\;\;\;\;\;\;\nn
\\\nn
\\
\left.+\int_{s_L}^{s_R}iT\sqrt{q_-}ds+\frac{1}{2}\int_{s'_L}^{s''}i\omega
  ds-\frac{1}{2}\int_{s''}^{s'_R}i\omega  ds
-\int_{s'}^{s''}i\omega ds\right)
\label{5.14}
\end{eqnarray}    

Therefore, the final $exact$ formula for the transition amplitude  is:
\begin{eqnarray}
a_-(+\infty,T)=\frac{3a_La_R}{2T\Delta\epsilon}e^{-\int_{s'}^{s''}i\omega(s,T)ds+i\gamma(T)}\chi_{\bar
  1\to2}(T)
\label{5.15}
\end{eqnarray}
and the probability $P$ reads:
\begin{eqnarray}
P=\frac{9a_L^2a_R^2}{4T^2(\Delta\epsilon)^2}e^{-2\int_{s'}^{s''}i\omega
  ds}|\chi_{\bar 1\to2}(T)|^2
\label{5.16}
\end{eqnarray}
where in the last two formulae point $s'$ is an arbitrary point on the real axis while 
point $s''$ being the one of line $abc$ of Fig. 1 is taken to lie simultaneously on the anti-Stokes line passing by point $s'$.

The adiabatic limit of the transition probability is therefore:
\begin{eqnarray}
P=\frac{9a_L^2a_R^2}{4T^2(\Delta\epsilon)^2}e^{-2\Re\left(iT\int_{s'}^{s''}\mu
    B_0(s)ds\right)}
\label{5.17}
\end{eqnarray}
where $s''$ is now an arbitrary point of the continuous Stokes line
passing by roots of $B_0(s)$
closest to the real axis.

\section*{6. The general case of algebraic magnetic field}

\hskip +2em The result given by the formula (\ref{5.15}) can be easily generalized. From the way of 
obtaining 
formula (\ref{5.9}) it follows that the most important is the existence of the continuous Stokes line
$abcde$ on Fig. 2 and its $T\to+\infty$-limit, i.e. the Stokes line $abc$ of Fig. 1, which link the
respective infinities $\Re s=-\infty$ and $\Re=+\infty$  on both Stokes graphs. Another important
property was the way field ${\bf B}$ approached the limits ${\bf
  B}^{\pm}$  when $\Re t\to\pm\infty$ respectively in the strip
${\bf\Sigma}$ mentioned in the assumption ${\bf 3}^0$. Let us therefore accept the following two additional
assumptions:

     ${\bf 7}^0$ There are two Stokes lines on each of the Stokes
     graphs corresponding to $iT\sqrt{q_\pm}$   
     which can be taken as the boundaries of the strip ${\bf \Sigma}$. Each of these two Stokes lines links
     continuously both infinities of the strip ${\bf \Sigma}$, see Fig.3;
     
${\bf 8}^0$ Inside the strip ${\bf \Sigma}$ the field ${\bf B}$ approaches the infinities of the strip according to the
     following asymptotic formula:
\begin{eqnarray}
{\bf B}(sT,T)\sim {\bf B}_0^{\pm}(T)+\frac{{\bf
    B}_1^{\pm}(T)}{s^{\alpha_1^{\pm}}}+\frac{{\bf
      B}_2^{\pm}(T)}{s^{\alpha_2^{\pm}}}+\ldots+\frac{{\bf
        B}_k^{\pm}(T)}{s^{\alpha_k^{\pm}}}+\ldots,\;\;\;\;\;\;\Re
    s\to\pm\infty\nn\\
\label{6.1}
\\
\frac{1}{2}<\alpha_1^{\pm}<\alpha_2^{\pm}<\ldots<\alpha_k^{\pm}<\ldots\hspace{50mm}\nn
\end{eqnarray}
where $\alpha_1^{\pm},\ldots,\alpha_k^{\pm}$, are rational if ${\bf B}^2$ is a
meromorphic function of $s$.

     If the Stokes graph corresponding to $iT\sqrt{q_-}$  satisfies the conditions of being a graph
of the ALR-system described in Sec.4, then we can claim that there are
four sectors $S_1,S_{\bar 1},S_2,S_{\bar 2}$
 of the graph and the corresponding fundamental solutions
 $\chi_1,\chi_{\bar 1}$ which can be
used in exactly the same way as it was done in the case of the Nikitin model to solve the problem
stated in Sec.2, see Fig.4.
\vskip 12pt
\begin{tabular}{cc}
\psfig{figure=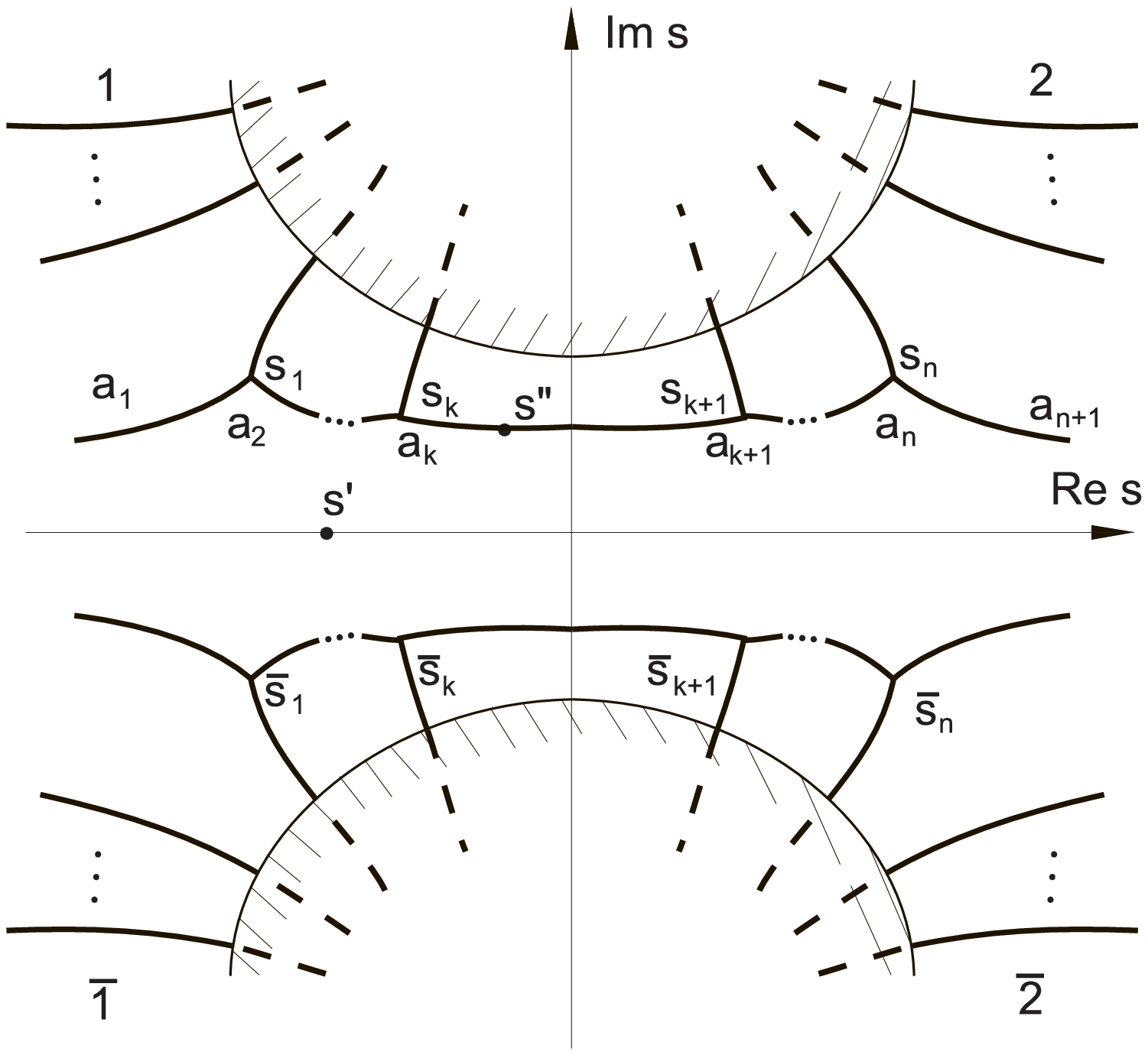,width=7cm}& \psfig{figure=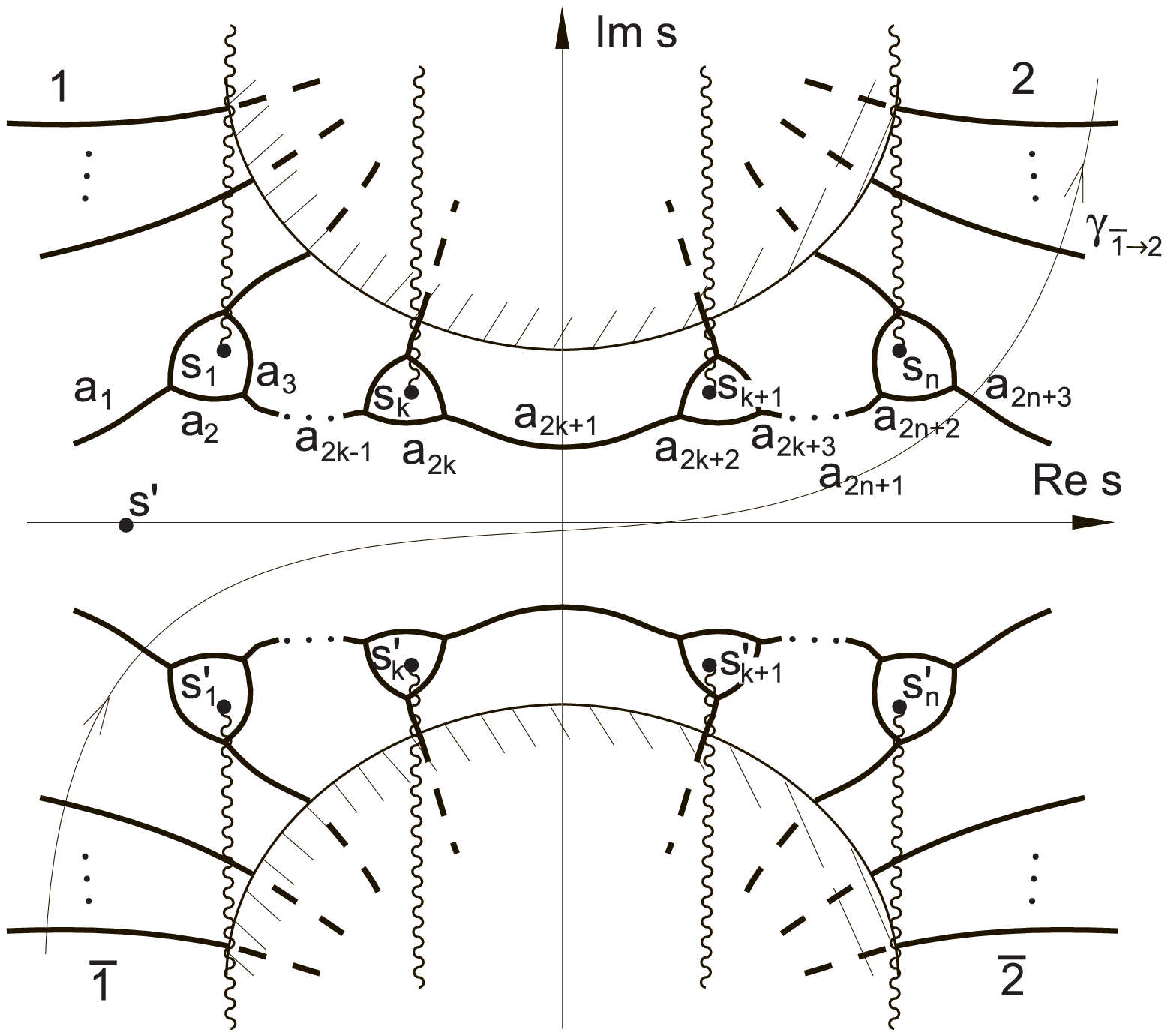,width=7cm}\\
Fig.3 The Stokes graph corresponding& Fig.4 The Stokes graph
corresponding\\
to general $q^{(0)}(s)$ considered in Sec.6&to general $q_-(s,T)$
considered in Sec.6
\end{tabular}
\vskip 12pt

     Let us choose the $xyz$-axes in the space of vector ${\bf B}$ in such a way that one of its limit
components $B_{x,0}^{\pm}$ and $B_{y,0}^{\pm}$ does not vanish in the corresponding infinities. Let us also assume that
vectors ${\bf B}_0^\pm(T)$ and ${\bf B}_1^\pm(T)$ of expansion (\ref{6.1}) are not parallel to each other in the respective
infinities (otherwise we should take another pair of vectors appearing in (\ref{6.1}) satisfying the last
property and having the smallest sum of the power exponents by which they are accompanied).
Then, if we take into account the following asymptotic which comes out
of (\ref{2.6}) and of the
above assumptions when $\Re s\to{\pm}\infty$ inside the strip:

\begin{eqnarray}
c\sim\left(-\frac{1}{2}\frac{\left[{\bf B}_0^\pm\times\left({\bf
          B}_0^\pm\times{\bf B}_1^\pm\right)\right]_z}{B_0^{\pm
        2}\sqrt{B_{x,0}^{\pm 2}+B_{y,0}^{\pm 2}}}+
\frac{i}{2}\frac{\left({\bf B}_0^\pm\times{\bf B}_1^\pm\right)_z}{B_0^\pm\sqrt{B_{x,0}^{\pm 2}+B_{y,0}^{\pm 2}}}\right)\frac{1}{s^{\alpha_1^\pm}}\equiv\frac{D^\pm}{s^{\alpha_1^\pm}}\nn 
\\\nn
\\
\omega\sim\mu TB_0^\pm+\left(\mu T{\bf B}_0^\pm\cdot{\bf
    B}_1^\pm-\frac{B_{z,0}^\pm}{B_0^\pm}\frac{\left({\bf B}_0^\pm\times{\bf
      B}_1^\pm\right)_z}{\sqrt{B_{x,0}^{\pm 2}+B_{y,0}^{\pm 2}}}\right)\frac{1}{s^{\alpha_1^\pm}}\equiv\frac{G^{\pm}}{s^{\alpha_1^\pm}}\nn
\\\nn
\\
\frac{1}{2}\left(\frac{{\dot
          c}^*}{c^*}-i\omega\right)+iT\sqrt{q_-}\sim\left\{
\begin{array}{c}-i\mu
  TB_0^-\frac{\alpha_1^+}{s}\\\frac{D^-(D^-)^*}{i\mu
          TB^-_0}\frac{1}{s^{2\alpha_1^-}}
\end{array}
\right.\hspace{20mm}
\label{6.2}
\\\nn
\\
\frac{1}{2}\left(\frac{{\dot c}^*}{c^*}-i\omega\right)-iT\sqrt{q_-}\sim\left\{\begin{array}{c}-\frac{D^+(D^+)^*}{i\mu
  TB^+_0}\frac{1}{s^{2\alpha_1^+}}\\-i\mu
  TB^-_0\frac{\alpha_1^-}{s}
\end{array}
\right.\hspace{18mm}\nn
\\\nn
\\
\frac{\dot c}{c}\sim-\frac{\alpha_1^\pm}{s}\hspace{50mm}\nn
\end{eqnarray}
then we can repeat the procedure of
the previous section to get the analogues of formulas (\ref{5.9})
and (\ref{5.14}). Namely, we have for them:
\begin{eqnarray}
a_-(+\infty,T)=\frac{1}{\mu T\sqrt{B_0^-(T)B^+_0(T)}}\exp\left(\int_{-\infty}^{s_0}\left[-\frac{1}{2}\left(\frac{\dot
  c}{c}-i\omega\right)+iT\sqrt{q_-}\right]ds\right.+\nn
\\\nn
\\
\left.\ln c(s_0)-\int_{s'}^{s_0}i\omega
  ds+\int_{s_0}^{+\infty}\left[\frac{1}{2}\left(\frac{\dot
  c}{c}-i\omega\right)+iT\sqrt{q_-}\right]d\sigma\right)\chi_{\bar 1\to2}(T)=\\
\nn\\
\frac{a_La_R}{\mu
  T}\sqrt{\frac{|D^-(T)D^+(T)|}{B^-_0(T)B^+_0(T)}}e^{-\int^{s''}_{s'}i\omega(s,T)ds+i\gamma}\chi_{\bar 1\to2}(T)\hspace{15mm}\nn
\label{6.3}
\end{eqnarray}
where points $s'$ and $s''$ have been chosen again on the same
anti-Stokes line of the graph corresponding to $i\omega(s,T)$ and
\begin{eqnarray}
P(T)=\frac{a_L^2a_R^2|D^-(T)D^+(T)|}{(\mu
  T)^2B_0^-(T)B^+_0(T)}e^{-2\Re\int_{s'}^{s''}i\omega(s,T)ds}|\chi_{\bar
  1\to2}(T)|^2
\label{6.4}
\end{eqnarray}
where $D^\pm$ are given by:
\begin{eqnarray}
D^{\pm}=-\frac{1}{2}\frac{\left[{\bf B}^\pm_0\times\left({\bf
      B}^\pm_0\times{\bf
      B}^\pm_1\right)\right]_z}{B^{\pm2}_0\sqrt{B^{\pm2}_{x,0}+B^{\pm2}_{y,0}}}+\frac{i}{2}\frac{({\bf
      B}_0^\pm\times{\bf B}^\pm_1)_z}{B^{\pm}_0\sqrt{B^{\pm2}_{x,0}+B^{\pm2}_{y,0}}}
\label{6.5}
\end{eqnarray}
so that:
\begin{eqnarray}
|D^\pm|=\frac{B^\pm_1\sin\phi^\pm}{2B_0^\pm}
\label{6.6}
\end{eqnarray}
where $\phi^\pm(T)$ are the angles between fields ${\bf
  B}^\pm_0$ and ${\bf B}^\pm_1$ respectively.

Again, the exact form of the coefficients $a_{L,R}$ can be found if the
exact equations of the Stokes lines corresponding to $\omega(s,T)$ and
$q_-(s,T)$ are known.

Therefore, the final forms of the transition probability and its adiabatic
limit are:
\begin{eqnarray}
P(T)=\frac{a_L^2a_R^2B^-_1(T)B^+_1(T)\sin\phi^-(T)\sin\phi^+(T)}{\left(2\mu
    TB^-_0(T)B^+_0(T)\right)^2}\;e^{-2\Re\int^{s''}_{s'}i\omega(s,T)ds}|\chi_{\bar1\to2}(T)|^2
\label{6.7}
\end{eqnarray}
and:
\begin{eqnarray}
P^{ad}=\frac{a_L^2a_R^2B^-_{1,0}B^+_{1,0}\sin{\phi_0^-}\sin{\phi_0^+}}{\left(2\mu TB^-_{0,0}B^+_{0,0}\right)^2}\;e^{-2\Re\left(iT\int^{s''}_{s'}\mu
    B_0(s)ds\right)}
\label{6.8}
\end{eqnarray}
where to get the last formula the asymptotic expansion (\ref{2.1}) has been applied to fields ${\bf
B}^\pm_0(T)$ and ${\bf B}^\pm_1(T)$ as well as to $\omega$ given by
(\ref{2.6}). Point $s''$ is now an arbitrary point of the continuous
Stokes line $a_1a_2\ldots a_na_{n+1}$ passing by the roots of $B_0(s)$
closest to the real axis, as it is shown on Fig.3. Note that because of
our assumption the angles in (\ref{6.7}) and (\ref{6.8}) are different
from $0$ and $\pi$.

\section*{7. Another two examples with exponentially decreasing magnetic fields}

\hskip +2em   We consider here another two examples of magnetic fields depending exponentially
on time. The main difference between these cases and those considered in the previous sections lies in the number
of level crossings which in the exponential cases is, of course, infinite.

     We consider the following two cases of the fields:
\begin{eqnarray}
{\bf a})\;\;\;\;\;\;\;\;\;{\bf B}(t,T)={\bf B}_0+\frac{{\bf
     B}_1}{\cosh \left(\frac{t}{T}\right)},\;\;\;\;{\bf B}_0\cdot{\bf
     B}_1=0,\;\;\;\;\;B_0=|{\bf B}_0|\ne|{\bf B}_1|=B_1\nn
\\
\label{7.1}
\\\nn
{\bf b})\;\;\;\;\;\;\;\;\;{\bf B}(t,T)={\bf B}_0+{\bf
     B}_1}{\tanh \left(\frac{t}{T}\right),\;\;\;\;{\bf B}_0\cdot{\bf
     B}_1=0,\;\;\;\;\;|{\bf B}_0|={\bf B}_1|=B_0\nn
\end{eqnarray}
\vskip 20pt
$\underline{Case\; {\bf a})}$
\vskip 20pt
     The relevant quantities for this case have the forms:
\begin{eqnarray}
c(s,T)=\frac{1}{2}\frac{B_0B_1\sinh s}{B_1^2+B_0^2\cosh^2 s}\hspace{30mm}\nn\\
\nn\\
\omega(s,T)=\mu T\sqrt{B_0^2+\frac{B_1^2}{\cosh^2s}}\hspace{30mm}\nn\\
\nn\\
q_-(s,T)=-\frac{1}{4T^2}\left(\coth
  s-\frac{B_0^2\sinh(2s)}{B_1^2+B_0^2\cosh^2 s}-i\mu
  T\sqrt{B_0^2+\frac{B_1^2}{\cosh^2s}}\right)^2+
\label{7.2}
\nn\\
\\
\frac{i\mu}{2T}\frac{B_1^2\sinh
  2s}{\sinh^2s\left(B_1^2+B_0^2\cosh^2s\right)^\frac{1}{2}}+\frac{1}{4T^2}\frac{B_0^2B_1^2\sinh^2s}{\left(
  B_1^2+B_0^2\cosh^2 s\right)^2}-\nn\\
\nn\\
\frac{1}{2T^2}\left(\frac{1}{\sinh^2s}+\frac{2B_0^2\cosh2s}{B_1^2+B_0^2\cosh^2s}-\frac{B_0^4\sinh^2
   2s}{\left(B_1^2+B_0^2\cosh^2s\right)^2}\right)\nn
\end{eqnarray}
and the Stokes graphs defined by $\omega(s,T)$ and $q_-(s,T)$ are
shown on figures 5. and 6. respectively. 

\vskip 12pt
\begin{tabular}{cc}
\psfig{figure=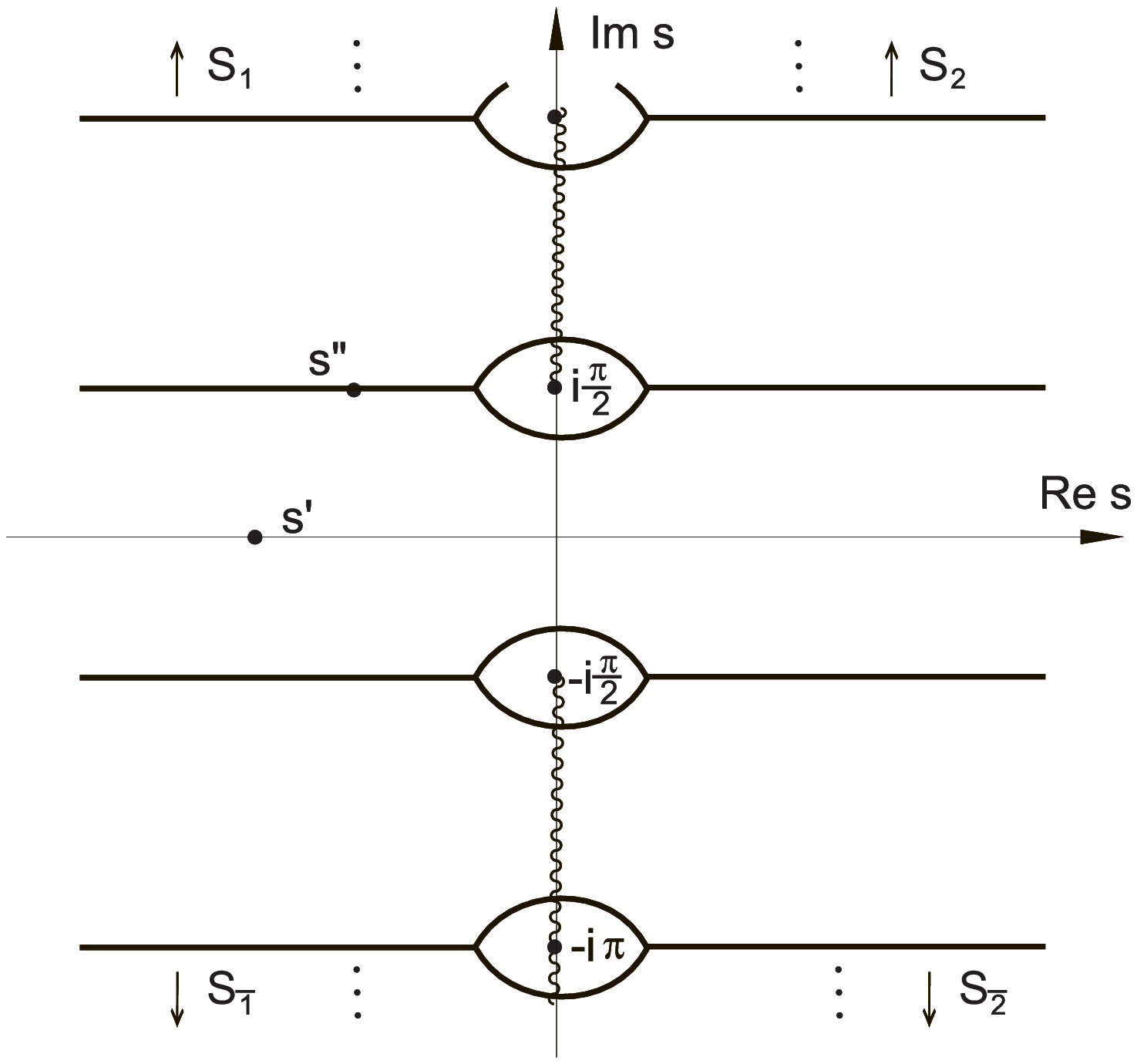,width=7cm}& \psfig{figure=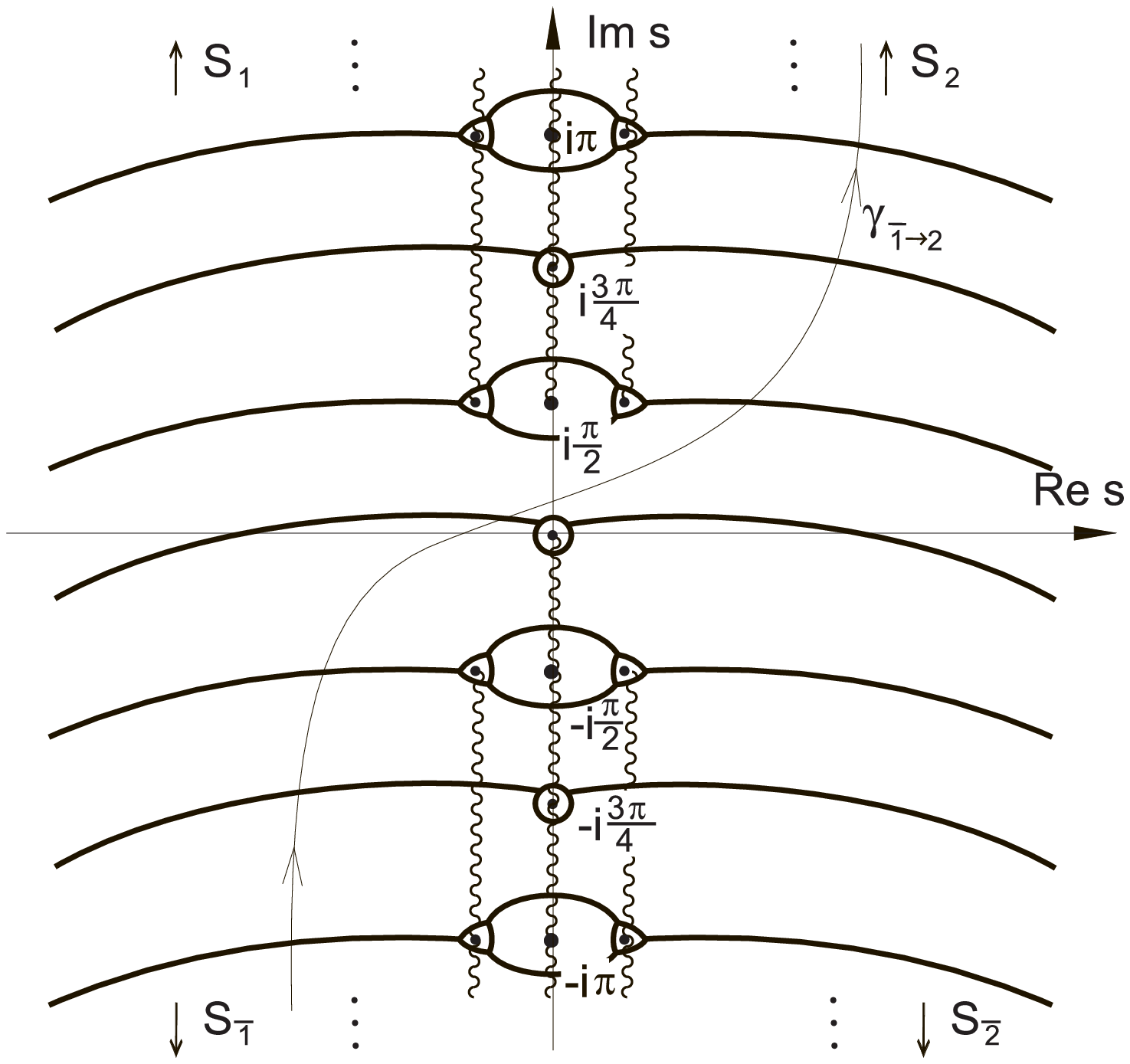,width=7cm}\\
Fig.5 The Stokes graph corresponding& Fig.6 The Stokes graph
corresponding\\
to $q^{(0)}(s)$ of case {\bf a}) of (\ref{7.1})&to $q_-(s,T)$ of (\ref{7.2})
\end{tabular}
\vskip 12pt

The procedure leading us to formula (\ref{5.9}) is still valid but the corresponding
sectors $S_1,\;S_{\bar 1},\;S_2,\;S_{\bar 2}$ are now less exposed. Namely, the first two lie on the left of the
imaginary axis, $S_1$ above and $S_{\bar 1}$ below the real axis whereas the next two on lie the right of the
imaginary axis and, respectively, above and below the real axis. A peculiarity of this and the
next case is that these sectors are cut by the infinite number of the Stokes lines parallel to the
real axis and distributed up and down to the imaginary infinities, see figures 3. and 4.. The
fundamental solutions defined in these sectors vanish in their imaginary infinities. Therefore, the
corresponding transition amplitude $a_-(s,T)$ from level $E_+$ to
$E_-$ looks as follows:
\begin{eqnarray}
a_-(T)=\frac{2iB_1a_La_R}{\mu B_0
  \sqrt{1+\mu^2T^2B_0^2}}e^{-\int_{s'}^{s''}i\omega(s,T)ds+i\gamma(T)}\chi_{\bar 1\to 2}(T)
\label{7.3}
\end{eqnarray}

To get the above formula we have taken into account the following asymptotic behaviour of the quantities determining
it:
\begin{eqnarray}
c(x+iy)\sim\left\{\matrix{+\frac{B_1}{B_0}e^{-x-iy}\;\;\;\;x\to
  +\infty\nn
\\-\frac{B_1}{B_0}e^{+x+iy}\;\;\;\;x\to -\infty}\right.
\;\;\;\;\;\;\;\;\;\frac{\dot
  c(s)}{c(s)}\sim\left\{\matrix{-1\;\;\;\;\Re s\to+\infty\nn
\\+1\;\;\;\;\Re s\to-\infty}\right.\nn\\
\\
\omega(x+iy,T)\sim-\mu TB_0,\;\;\;\;|x|\to\infty,\;\;\;\;\;\;y'_{L,R}-y_{L,R}\sim
\frac{2\ln a_{L,R}}{\mu TB_0}+\frac{|x_{L,R}|}{\mu
  TB_0},\;\;\;\;\;|x_{L,R}|\to\infty\nn
\label{7.4}
\end{eqnarray}
where $s_{L,R}=x_{L,R}+iy_{L,R}$ and $s'_{L,R}=x'_{L,R}+iy'_{L,R}$ have the same meaning as previously, i.e. lie on the
corresponding Stokes lines defined by  $q_-(s,T)$ and $\omega(s,T)$, respectively, whilst $a_{L,R}$ measure
(together with the terms linear in $x_{L,R}$) the deviations of these lines at the corresponding infinities.

     Therefore, for the exact transition probability and its adiabatic limit, we obtain from (\ref{7.3}):
\begin{eqnarray}
P(T)=\frac{\left(2B_1a_La_R\right)^2}{\mu^2 B_0^2
  \left(1+\mu^2T^2B_0^2\right)^2}e^{-2\Re\int_{s'}^{s''}i\omega(s,T)ds}|\chi_{\bar 1\to 2}(T)|^2
\label{7.5}
\end{eqnarray}
and
\begin{eqnarray}
P^{ad}=\left(\frac{2B_1a_La_R}{\mu^2TB_0^2}\right)^2\exp\left(-2\mu
T\Re\int_{s'}^{s''}i\sqrt{B_0^2+\frac{B_1^2}{\cosh^2s}}ds\right)
\label{7.6}
\end{eqnarray}
respectively.
\vskip 20pt
$\underline{Case\; {\bf b})}$
\vskip 20pt
In this case we have:
\begin{eqnarray}
c(s,T)=-\frac{1}{2\cosh(2s)},\hspace{30mm}\frac{\dot
  c(s,T)}{c(s,T)}=-2\tanh(2s)\nn\\
\nn\\
\omega(s,T)=\mu
  TB_0\frac{\sqrt{\cosh(2s)}}{\cosh s}\hspace{30mm}\nn\\
\label{7.7}\\
q_-(s,T)=-\frac{1}{4T^2}\left(2\tanh(2s)+i\mu
  TB_0\frac{\sqrt{\cosh(2s)}}{\cosh
    s}\right)^2-\nn\\
\nn\\
\frac{i\mu B_0}{2T}\frac{\tanh (2s)-\tanh s}{\cosh s}\sqrt{\cosh(2s)}-\frac{7}{4T^2}\frac{1}{\cosh^2(2s)}\nn
\end{eqnarray}
and the Stokes graphs corresponding to $\omega(s,T)$ and $q_-(s,T)$ are shown in figures 7. and 8.
respectively.

\vskip 12pt
\begin{tabular}{cc}
\psfig{figure=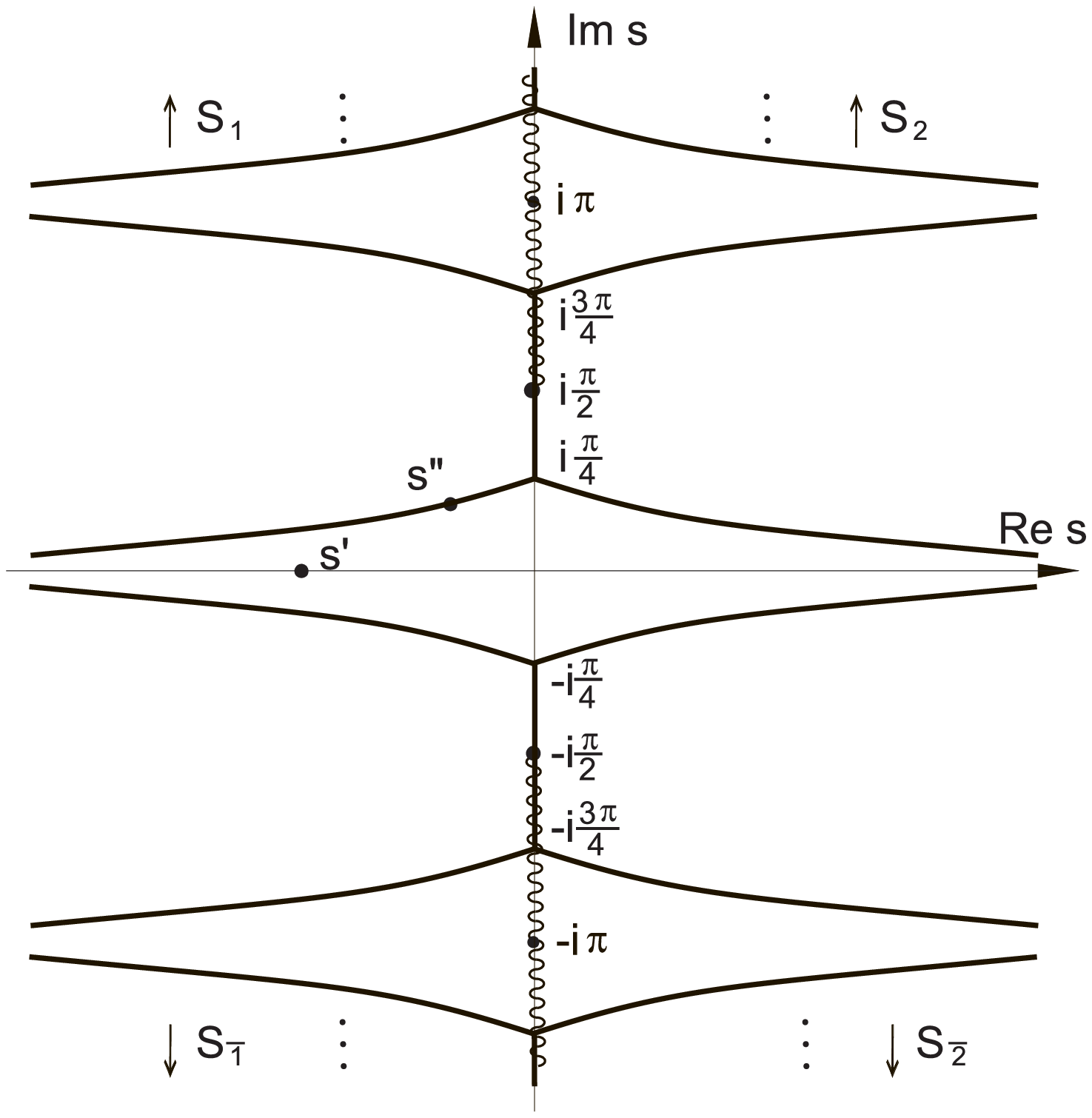,width=7cm}& \psfig{figure=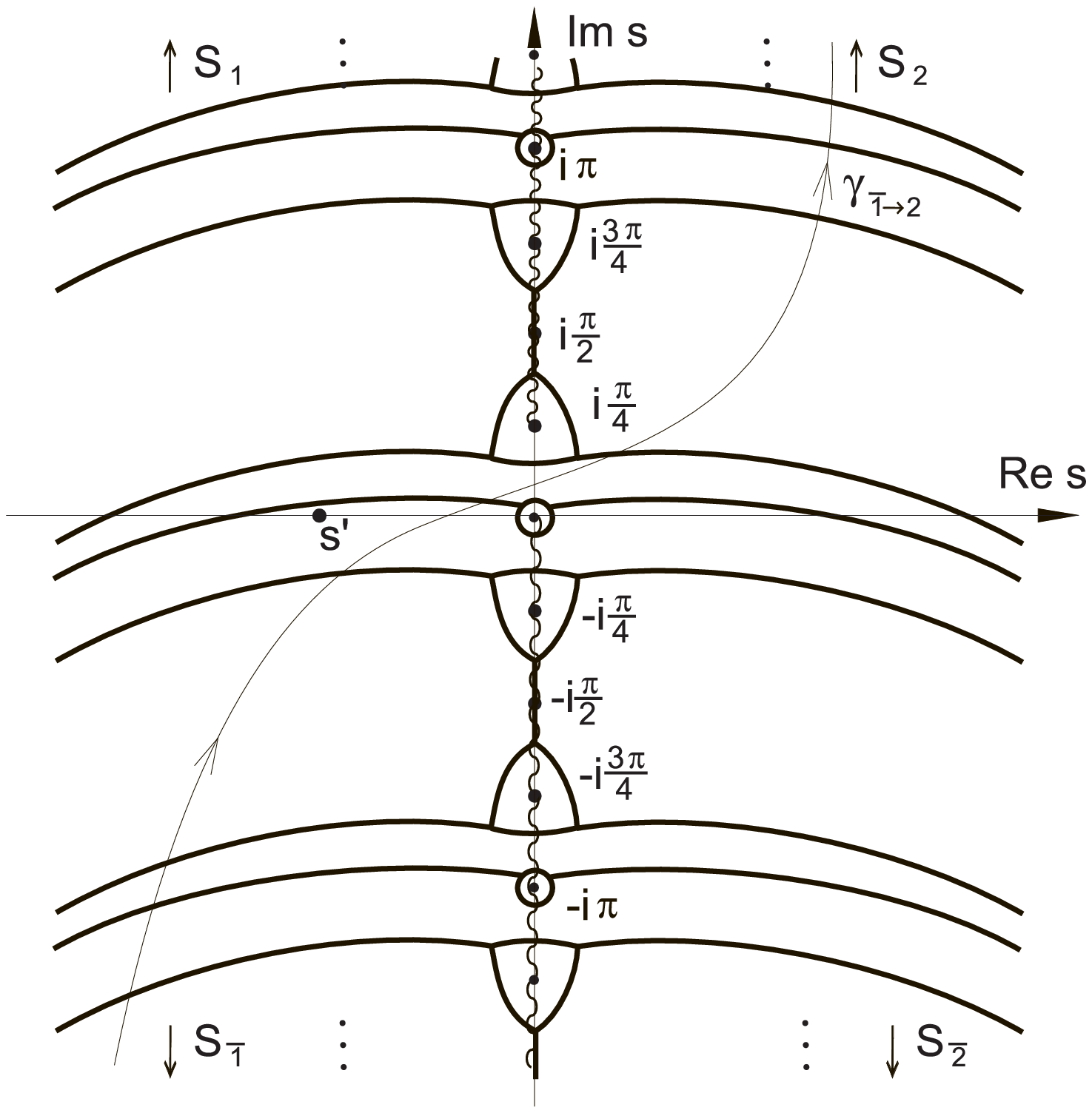,width=7cm}\\
Fig.7 The Stokes graph corresponding& Fig.8 The Stokes graph
corresponding\\
to $q^{(0)}(s)$ of case {\bf b}) of (\ref{7.1})&to $q_-(s,T)$ of (\ref{7.7})
\end{tabular}
\vskip 12pt

Again, the transition amplitude can be calculated taking into account the following
asymptotic:
\begin{eqnarray}
c(x+iy)\sim\left\{\matrix{-e^{-2x-2iy}\;\;\;\;x\to
  +\infty\nn
\\-e^{+2x+2iy}\;\;\;\;x\to -\infty}\right.
\;\;\;\;\;\;\;\;\;\frac{\dot
  c(s)}{c(s)}\sim\left\{\matrix{-2\;\;\;\;\Re s\to+\infty\nn
\\+2;\;\;\;\Re s\to-\infty}\right.\nn
\\\nn
\\
\omega(x+iy,T)\sim-\sqrt 2\mu TB_0,\;\;\;\;|x|\to\infty,\;\;\;\;\;\;\;\;\;\;\;\;\;\;\;\;
\label{7.8}
\\\nn
\\
y'_{L,R}-y_{L,R}\sim
\frac{2\ln a_{L,R}}{\sqrt 2\mu TB_0}+\frac{|x_{L,R}|}{\sqrt2\mu
  TB_0},\;\;\;\;\;|x_{L,R}|\to\infty\;\;\;\;\;\;\;\;\;\;\nn
\end{eqnarray}
so that we get for it:
\begin{eqnarray}
a_-(T)=\frac{2a_La_R}{\mu \sqrt{4+2\mu^2T^2B_0^2}}e^{-\int_{s'}^{s''}i\omega(s,T)ds+i\gamma(T)}\chi_{\bar 1\to 2}(T)
\label{7.9}
\end{eqnarray}

     Therefore, for the corresponding transition probabilities we obtain:
\begin{eqnarray}
P(T)=\frac{\left(2a_La_R\right)^2}{\mu^2 
  \left(4+2\mu^2T^2B_0^2\right)}e^{-2\Re\int_{s'}^{s''}i\omega(s,T)ds}|\chi_{\bar 1\to 2}(T)|^2
\label{7.10}
\end{eqnarray}
and
\begin{eqnarray}
P^{ad}=\left(\frac{\sqrt2a_La_R}{\mu^2TB_0}\right)^2\exp\left(-2\mu
TB_0\Re\int_{s'}^{s''}i\sqrt{\frac{\cosh(2s)}{\cosh s}}ds\right)
\label{7.11}
\end{eqnarray}
\vskip 20pt
\section*{8. Non vanishing contribution of the Berry phase}

\hskip +2em The previous sections have provided us with the examples
of Hamiltonians in which the corresponding transition probabilities
have had no contributions from the term $-\frac{B_z}{B}\frac{\left({\bf
      B}\times{\bf\dot B}\right)_z}{B_x^2+B_y^2}$ of $\omega$ (see
  (\ref{2.6})) representing (at least) a part of the Berry phase of
  the transition amplitudes. The Hamiltonian defined by the following
  field:
\begin{eqnarray}
{\bf B}=\frac{B_0}{1+s^2}\left[1,\alpha s,s^2\right],\;\;\;\;\alpha
>\sqrt{2}
\label{8.1}
\end{eqnarray}
provides us with the corresponding positive example of such a
contribution. Namely, for this case we have:
\begin{eqnarray}
\omega=\mu
TB_0\frac{\sqrt{\left(1+s^2\right)^2+\alpha^2-2}}{1+s^2}-\frac{1}{\sqrt{\left(1+s^2\right)^2+\alpha^2-2}}\frac{s^2}{1+\alpha^2s^2}
\label{8.2}
\end{eqnarray}

From (\ref{8.1}) it follows easily that for this case the transition probability (\ref{6.7}) takes the form
\begin{eqnarray}
P(T)=\frac{a_L^2a_R^2}{\left(2\mu
    TB_0\right)^2}\;e^{-2\Re\int^{s''}_{s'}i\omega(s,T)ds}|\chi_{\bar1\to2}(T)|^2
\label{8.3}
\end{eqnarray}

It is the second term of (\ref{8.2}) which is responsible for the Berry phase contribution
to the transition probability (\ref{8.3}).
We shall calculate this contribution in the adiabatic limit only and
for $\alpha$ close to $\sqrt{2}$. This
assumption allows us to calculate the corresponding path integral:
\begin{eqnarray}
{\bf I}_\gamma=-i\int_{\gamma}\frac{1}{\sqrt{\left(1+s^2\right)^2+\alpha^2-2}}\frac{s^2}{1+\alpha^2s^2}ds
\label{8.4}
\end{eqnarray}
 from
point $s=0$ to the closest root $s_0=i\sqrt{1+i\sqrt{\alpha^2-2}}$ of the polynomial
$\left(1+s^2\right)^2+\alpha^2-2$, lying in the second quadrant of the
$s$-plane. For $\alpha$ close to $\sqrt{2}$ we can simplify the
integration expanding suitably the square root in the integrand of
(\ref{8.4}) and the root $s_0$ as well. It is easy to check that under the
above assumptions the net result of such calculations is:
\begin{eqnarray}
-2\Re{\bf I}_\gamma=\ln \frac{\sqrt{\alpha-\sqrt{2}}}{2^\frac{1}{4}\left(\sqrt{2}-1\right)^{\sqrt{2}}}+O\left(\sqrt{\alpha-\sqrt{2}}\right)
\label{8.5}
\end{eqnarray} 

Obviously, the above Berry phase contribution to the transition
probability (\ref{8.3}) modifies its prexponential factor multiplying
it by the following
additional one:
\begin{eqnarray}
C=\frac{\sqrt{\alpha-\sqrt{2}}}{2^\frac{1}{4}\left(\sqrt{2}-1\right)^{\sqrt{2}}}
\label{8.6}
\end{eqnarray} 

\section*{9. Conclusions and discussion}

\hskip +2em     We have shown in this paper that the fundamental solution method has turned out to be
very effective also in its application to the problems of the transition amplitudes in two energy
level systems. In particular it has enabled us to obtain compact and exact formulae for these
amplitudes and to get easily their adiabatic approximations as well. Due the clear way
of their obtaining and their compact forms, the formulae allow us to claim that there are no particular effects coming out of the many complex level crossings i.e. there are no 
of individual contributions of any kind to the transition amplitude from each such crossing leading to any
particular interference effects in these amplitudes. Just the opposite, such a contribution is controlled
totally by the Stokes line closest to the real axis of the $t$-plane which is however built by these
crossing points of the two energy levels. This result is independent of both the number of
complex level crossings (i.e. finite or infinite) and of the particular type of the $t$-dependence of the
effective magnetic field (i.e. algebraic or exponential). In this way
the respective results of Joye, Mileti and Pfister [4] have not been
confirmed by our approach. 
This last
difference seems to be rather dramatic and, as it seems to us, its
origin lies in an erroneous calculation of the transition amplitude by
the authors mentioned. Namely, it is the formula (6.21) of their paper
[4] for the transition matrix $X(z_1)$ which is wrong, particularly if applied
further in their {\it Lemma 6.1} to get the general
formula of it. This can be seen if we rewrite our results in
terms of the transition matrix. 

Namely, let us denote the r.h.s. of
the formula (28) by $U_{21}(s,T)$ and the result we
obtain calculating $a_+(s,T)$ with the help of the second of Eqs.(5)
and of (28) by $U_{11}(s,T)$ . 

Reversing the problem we have solved in our paper by
assuming that for $s=-\infty$ the vanishing amplitude is rather
$a_+(s,T)$ than $a_-(s,T)$ we obtain by exactly the same methods as
used in Sec.3 and the further ones the remaining elements
$U_{12}(s,T)$ and $U_{22}(s,T)$ which construct the transition matrix
${\bf U}(s,T)$. For the choice we have done in our paper we have of
course:
\begin{eqnarray}
{\bf a}(s,T)\equiv\left[\begin{array}{c}a_+(s,T)\\a_-(s,T)\end{array}\right]=\left[\begin{array}{cc}U_{11}(s,T)&U_{12}(s,T)\\U_{21}(s,T)&U_{22}(s,T)\end{array}\right]\left[\begin{array}{c}1\\0\end{array}\right]
\label{9.1}
\end{eqnarray}

Matrix ${\bf U}(s,T)$ is of course unitary (for real $s$) and ${\bf
  U}(-\infty,T)={\bf I}$.

It should be clear that the order of $U_{12}(s,T)$ as a function of
     its arguments is the {\it same} as that of $U_{21}(s,T)$, the latter element
     being given by the r.h.s. of (28) so that the adiabatic limit of
     ${\bf U}(s,T)$ is given by:
\begin{eqnarray}
{\bf U}^{ad}(s,T)=\left[\begin{array}{cc}1&U_{12}^{ad}(s,T)\\U_{21}^{ad}(s,T)&1\end{array}\right]
\label{9.2}
\end{eqnarray}

 Therefore this matrix is {\it
     not} a triangular one in this limit, as it is the case of $X(z_1)$
     mentioned earlier, which does {\it not} contain the {\it non vanishing}
     element $U_{12}^{ad}(s,T)$. 

Moreover we can not apply matrix ${\bf
     U}(s,T)$ directly to continue the solution (\ref{9.1}) along the
     central strip of the corresponding Stokes graph from point
     $s$ to another one $s'$. The proper continuation is of course the
     following:
\begin{eqnarray}
{\bf a}(s',T)={\bf U}(s',T){\bf U}^{-1}(s,T){\bf a}(s,T)
\label{9.3}
\end{eqnarray}

In particular, if it is possible to continue the solution ${\bf
     a}(s,T)$ along, say, the upper Stokes line limiting the central strip
     (i.e the level crossings $s_1,\;s_2,\;...\;,s_n$ met along this line are not an obstacle to
     such a continuation) then continuing ${\bf
     a}(s,T)$ in this way to $s=+\infty$ we get:
\begin{eqnarray}
{\bf a}(+\infty,T)={\bf U}(+\infty,T){\bf U}^{-1}(s_n,T){\bf
  U}(s_n,T){\bf U}^{-1}(s_{n-1},T)\ldots\nn\\
\label{9.4}\\\nn
{\bf
  U}(s_2,T){\bf U}^{-1}(s_1,T){\bf
  U}(s_1,T)\left[\begin{array}{c}1\\0\end{array}\right]= 
{\bf
  U}(+\infty,T)\left[\begin{array}{c}1\\0\end{array}\right]\nn
\end{eqnarray}

The above results show that {\it none} of the contributions from the
     {\it individual} level crossings lying on the considered Stokes line survive on
     the way of continuation.

On the other hand writing both formula (6.21) and the respective result of {\it Lemma 6.1} of 
     [4] in terms of the quantities introduced above, we get:
\begin{eqnarray}
X(s_1)=\left[\begin{array}{cc}1&0\\U_{21}^{ad}(s_1,T)&1\end{array}\right]
\label{9.5}
\end{eqnarray}
and
\begin{eqnarray}
{\bf a}^{ad}(+\infty,T)=X(s_n)X(s_{n-1})\ldots X(s_2)X(s_1)\left[\begin{array}{c}1\\0\end{array}\right]
\label{9.6}
\end{eqnarray}

Comparing the last two formulas with the respective (\ref{9.2}) and
(\ref{9.4}) ones we see that formulas (\ref{9.5}) and (\ref{9.6}) are wrong. Particularly,
it is the incorrect formula (\ref{9.6}) which gives rise to the interference
effects in the amplitudes of Joye, Mileti and Pfister [4].

     Finally, we would like to mention that, as we have shown this in
     Sec.8 there are contributions to
     the transition probabilities originating from the geometrical
     (Berry) phase [16] although their geometrical meaning in the
     context of the transition amplitudes is not clear.

\section*{Acknowledgments}

\hskip +2em     We are greatly indebted to our colleagues, professors Piotr
     Kosi\' nski and Pawe{\l} Ma\'slanka,
for many discussions while working on the paper.

     This paper has been written under the support of the KBN grant no. 2P03B13416 (S.
Giller) and the \L\'od\'z University grant no. 795 (C. Gonera).

\end{document}